\documentclass[seceq]{ptptex}

\usepackage{graphicx}
\usepackage{wrapft}

\newcommand{\be}{\begin{equation}}
\newcommand{\ee}{\end{equation}}
\newcommand{\bea}{\begin{eqnarray}}
\newcommand{\eea}{\end{eqnarray}}

\title{
Modeling a Realistic Dynamical Model  \\ for High Energy Heavy Ion Collisions%
}

\author{
Chiho \textsc{Nonaka}$^{1,2}$\footnote{nonaka@hken.phys.nagoya-u.ac.jp} and Masayuki \textsc{Asakawa}$^{3}$\footnote{yuki@phys.sci.osaka-u.ac.jp} %
}

\inst{
$^1$Kobayashi-Maskawa Institute
for the Origin of Particles and the Universe (KMI), \\ $^2$Department of Physics, 
Nagoya University, \\ Furo-cho, Chikusa-ku, Nagoya 464-8602, Japan
\\
$^3$Physics Department, Osaka University, Toyonaka 560-0043, Japan 
}

\abst{
In this article, we outline the modeling of a realistic dynamical model
for comprehensive description of high energy heavy ion collisions. 
Comparing theoretical calculations
and experimental data at RHIC, we give detailed discussions on
the key ingredients for the construction 
of a multi-module model: initial condition, 
hydrodynamical expansion, hadronization, and freezeout processes. 
}

\begin{document}
\maketitle
\section{Introduction}
Since the Relativistic Heavy Ion Collider (RHIC) at Brookhaven National Laboratory (BNL) 
started its operation in 2000, a lot of discovery has been made and
a lot of insight related to quantum chromodynamics (QCD) 
phase transition and the Quark-Gluon Plasma (QGP) has been gained. 
One of the most physically interesting and
surprising outcomes at RHIC is the production of the 
strongly interacting QGP (sQGP). 
This accomplishment was realized by combining investigations from
both experimental \cite{white_papers} and theoretical \cite{QGP} sides. 
Because the QGP had been believed to be a weakly interacting system
like the ideal gas,
the discovery of sQGP opened up a new paradigm for the understanding of
high temperature and/or high density QCD. 

The highlights of the RHIC experiments were
(i) strong elliptic flow, which suggests early thermalization and 
early formation of collectivity; 
(ii) strong jet quenching, which confirms
the formation of hot and dense matter in collisions;
(iii) constituent quark number scaling of the elliptic flow, 
which indicates the formation of deconfined hot quark soup \cite{white_papers}. 
From studies of these experimental results with relativistic hydrodynamical 
models, jet energy loss mechanism, and recombination models,
the discovery of sQGP at RHIC \cite{QGP} was accomplished.
Furthermore, heavy ion collision operation at the Large Hadron Collider (LHC), 
whose collision energy is around 15 times as large as that at RHIC, started in 2010. 
Such a high energy collision experiment gives us an opportunity to perform further 
investigation on the QCD diagram. 

\begin{figure}
\centerline{\includegraphics[width=14 cm]{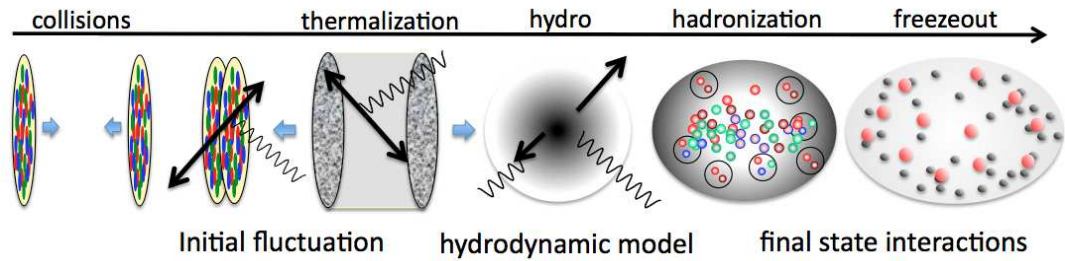}}
\caption{Schematic sketch of relativistic heavy ion collisions. The wavy lines 
stand for photons and the arrows stand for jets.  }
\label{fig:heavy-ion}
\end{figure}

Figure \ref{fig:heavy-ion} shows a schematic sketch of the time evolution in
relativistic heavy ion 
collisions on the basis of present understanding.  
When two heavy ions are accelerated to high energy,  
an extreme state which is described by static color charge and classical 
gluon fields, called the color glass condensate (CGC) state,
is realized inside each heavy ion. 
At the collision of the two heavy ions,
high energy prompt photons and Drell-Yan dileptons,
and jets are produced by hard scattering 
of quarks and gluons.  
After the collision thermalization is achieved in a short time.
During that time,
prethermal photons and dileptons are created and the entropy of the fireball increases. 
If the collision energy is large enough, QGP is also produced at this stage.  
Then hydrodynamical expansion starts, and a lot of interesting 
physical processes occur: collective flow formation, jet quenching, 
production of thermal photons and dileptons, and so on. 
As the temperature and density of the fireball decreases,
hadronization of the QGP phase takes place.
To understand the hadronization mechanism in relativistic 
heavy ion collisions, the recombination/coalescence mechanism and 
the fragmentation mechanism are employed.  
Gradually the mean free path between hadrons becomes large, and
eventually freezeout happens.
At this stage, final state interactions play an important role. 

The present understanding of heavy ion collisions strongly
suggests (Fig.~\ref{fig:heavy-ion}) 
that a multi-module modeling is indispensable for the description of entire history
of heavy ion collisions.
Knowledge of dominant physics at each stage has been accumulated,
but a comprehensive model is still missing. 
For the construction of such a multi-module model,
hydrodynamical models are a promising starting point,
because at present it is considered as one of the most reliable and 
successful dynamical models for understanding experimental
data at RHIC and LHC comprehensively, especially for the description
of the QGP phase.
At the same time, it is easy to implement the latest developments in
the physics of heavy ion collisions such as 
fluctuated initial conditions, the lattice QCD inspired equation of state, recombination mechanism
for hadronization, 
and final state interactions in freezeout processes into a hydrodynamical model, 
as we will describe later in detail.  

In Tabs. \ref{table:ideal-hydro} and \ref{table:viscous-hydro} 
current hydrodynamical models are listed. 
In line with the physical picture shown in Fig.~\ref{fig:heavy-ion},
in particular, we pick up and compare the following 
features of each hydrodynamical model: dimension of hydrodynamical expansion (dim), 
initial condition (IC), equation of state (EoS), and treatment of freezeout process. 
In addition, we also compare the numerical schemes 
in solving the relativistic
hydrodynamical equation and observables calculated in each model
in the tables. The importance of utilizing proper numerical schemes in
solving the relativistic hydrodynamical equation will be discussed in 
Sec. \ref{sec:NS}.
Recent development in relativistic viscous hydrodynamical models 
is remarkable. From the point of view of multi-module modeling, however,
the status of ideal hydrodynamical models is considered to be more mature. 
\begin{table}[htbp]
\caption{Ideal hydrodynamical models. In the table, we use the following abbreviation. 
IC: initial condition, G: Glauber model, CGC: color glass condensate, 
MC-G: Monte Carlo Glauber model, MC-CGC: Monte Carlo CGC,
lQCD: lattice QCD inspired EoS, SPH: smoothed particle hydrodynamics, 
PPM: piecewise parabolic method, 
CE: continuous emission, Obs: calculated observables, and
PD: particle distribution.}
\label{table:ideal-hydro}
\begin{tabular} {|c|c|c|c|c|c|c|}
\hline \hline
Ref. & dim & IC & EoS & scheme & freezeout & Obs\\
\hline 
Hama \cite{Hama:2004rr} &  3+1 & NeXus &  Bag model &  SPH  &  CE   & PD, $v_2$, HBT \\
\hline 
Hirano \cite{Hirano:2005xf} & 3+1 &G, CGC & Bag model  & PPM\footnotemark& cascade(JAM) &   $v_2$ \\
\hline 
Nonaka\cite{Nonaka:2006yn} & 3+1 & G & Bag model & Lagrange    &   cascade(UrQMD)   & PD, $v_2$ \\ 
\hline 
Hirano\cite{Hirano:2010jg, Hirano:2010je} &  3+1&  MC-G, MC-CGC & lQCD & PPM\footnotemark & cascade(JAM) & $v_2$\\ 
\hline 
Petersen\cite{Petersen:2008dd} & 3+1& UrQMD &  hadron gas  & SHASTA & cascade(UrQMD) & PD \\
\hline 
Holopainen \cite{Holopainen:2010gz} & 2+1 & MC-G & lQCD & SHASTA &  resonance decay&  $v_2$\\
\hline 
\end{tabular}
\end{table}
\footnotetext{
It is now customary in numerical hydrodynamics
not to call a hydrodynamical computer program a PPM
(piecewise parabolic method)\cite{CoWo84}
one unless it fulfills not only parabolic interpolation
of variables but also sharpening of discontinuity profiles and
flattening of post-shock oscillations\cite{Baiotti}.
Hirano's code\cite{Hirano:2005xf, Hirano:2010jg, Hirano:2010je} executes only the first of the above
three ingredients\cite{HiNa_PTEP}.
Attention needs to be payed on this difference when
comparing with other PPM codes
and estimating its capability to capture shocks.
}
\begin{table}[htbp]
\caption{Viscous hydrodynamical models. In the table, we use the following abbreviation. CD: central difference, and KT: Kurganov-Tadmor (KT) scheme.}
\label{table:viscous-hydro}
\begin{tabular} {|c|c|c|c|c|c|c|}
\hline \hline
Ref. & dim & IC & EoS & scheme & freezeout & Obs.\\ \hline 
Romatschke \cite{Romatschke:2007mq} & 2+1 & G & lQCD  & CD   &    single $T_{\rm f}$    & $v_2$ \\ \hline 
Dusling \cite{Dusling:2007gi} & 2+1 & G & ideal gas &      $-$          &   viscous correction        & $v_2$\\ \hline 
Luzum \cite{Luzum:2008cw} &  2+1     & G, CGC &      lQCD   &   CD       &  resonance decay        &   $v_2$    \\ \hline 
Schenke \cite{Schenke:2010rr} & 3+1 &   MC-G  &     lQCD         &    KT             &     viscous correction       &  $v_2$, $v_3$  \\ \hline 
Song \cite{Song:2010mg} & 2+1 & MC-G, MC-CGC & lQCD & SHASTA    & cascade(UrQMD) & $v_2$ \\ \hline
Chaudhuri \cite{Chaudhuri,Roy:2011pk}  & 2+1 &     G      & Bag model  & SHASTA &       viscous correction              &  $v_2$             \\ \hline 
Bozek \cite{Bozek:2011ua} & 3+1 &        G      &      lQCD          &  $-$       &       THERMINATOR2           &   $v_1$, $v_2$, HBT    \\ \hline 
\end{tabular}
\end{table}

In this article, we outline the modeling of a realistic dynamical model 
for the description of relativistic heavy ion collisions in line with the
physical picture shown by the schematic 
sketch (Fig.~\ref{fig:heavy-ion}). In the discussion, we refer to and 
compare with experimental data at SPS, RHIC, and LHC.  
The article is organized as follows. In Sec.\ \ref{sec:IC},
we review the initial condition for hydrodynamical models from the conventional 
Glauber type one to the latest attempt to include event-by-event fluctuated initial 
conditions. 
In Sec.\ \ref{sec:hydro}, we present the basic concepts and ingredients
of both ideal and viscous
hydrodynamics such as relativistic hydrodynamical equation,
equations of state, and transport coefficients. 
We will also discuss interplay among jets, medium, and hydrodynamical expansion. 
In Sec.\ \ref{sec:NS}, we review the numerical schemes 
which are listed in Tabs.~\ref{table:ideal-hydro} and \ref{table:viscous-hydro} and 
show the result of the newly developed scheme by one of the authors and her
collaborators for a relativistic viscous hydrodynamical model.  
In Sec.\ \ref{sec:hadronization},
we explain the recombination model and show its utility
in understanding hadron observables and the QCD phase diagram. 
In Sec.\ \ref{sec:freezeout},
we explain chemical and thermal freezeout processes
and discuss effects of final state interactions by
comparing theoretical calculation and experimental data. 
Section \ref{sec:sum} is devoted to summary and conclusions.

\section{Initial conditions \label{sec:IC}}   

The hydrodynamical equations of motion requires inputs of initial conditions for
all their dynamical variables, which are then evolved forward in
time.
These initial conditions are outside of the framework of hydrodynamical models 
and have to be determined by other means. 
Physically, they are
determined by the process during the initial collision of the nuclei
and the succeeding stage that makes the system approach to equilibrium,
which is eventually reached at a time 
$\tau_0$. Note that $\tau_0$, in principle, can depend on
the coordinate space rapidity $\eta$, while in practice it
is assumed to be independent of $\eta$.
The equilibration time is  
a parameter since the equilibration mechanisms are still 
under debate and the first principle determination of the initial conditions
of the equilibrated plasma phase has not been achieved.
\cite{Baier:2000sb,re-early-therm}.

Historically, parameterized initial conditions for entropy density
(or alternatively the energy densities) and the net baryon density have
been used \cite{Nonaka:2006yn,Kolb:2001qz,Hirano:2002ds,Teaney:2000cw}.  
In the transverse plane these distributions have been mainly parameterized 
based on Glauber-type models of nuclear collisions. In the longitudinal 
direction initial distributions inspired by Bjorken's scaling solution
are often used. Then a few parameters remain to be fixed additionally
in the initial condition, such as the maximum values of the energy or entropy density, and net 
baryon density. They are usually fixed by comparison with experimental data on
single particle rapidity distributions and transverse momentum spectra.

As a first trial, one can choose to set the initial 
longitudinal flow to Bjorken's scaling solution \cite{Bjorken:1982qr}, 
and one can set the initial transverse flow to zero. This simplest 
initial flow profile has served as the basis for all further investigation.
The possibility of the existence of an initial transverse flow at $\tau_0$
was discussed, e.g.,
by Kolb and Rapp \cite{Kolb:2002ve}. The initial flow improves the 
results for $P_T$-spectra and reduces the anisotropy. 
This suggests that HBT analyses may be a sensitive tool for the determination 
of the initial longitudinal flow. We note that, for the analysis of the final state
longitudinal flow, the Yano-Koonin parametrization is effective.
Hydrodynamical calculations during the early RHIC years did show 
serious disagreement with experimental data, especially for the ratio of 
$R_{\rm out}/R_{\rm side}$, leading to the notion of the HBT puzzle \cite{HeKo02}.
It turned out that the solution of the HBT puzzle is not so simple because it is
related to all stages of hydrodynamical model: initial conditions, the equation of state,
viscosity effect, and final state interactions.  
Pratt  \cite{Pratt:2008qv} showed that this HBT puzzle comes from not a single shortcoming of 
hydrodynamical models,  but the combination of several effects; it is solved by mainly
prethermal acceleration, a stiffer equation of state, and viscosity effect. 

Let us come back to the apparent early thermalization times found
at RHIC. Usually it is argued that small initial times $\tau_0$ are 
needed to describe the elliptic flow data as the elliptic flow builds up 
at the earliest stage of expansion when the eccentricity of the fireball is
largest \cite{Kolb:2000fha,Huovinen:2001cy}. 
However, we note that with suitable sets of initial conditions and 
freezeout temperatures in fact a larger initial proper time 
is also compatible with data. Luzum and Romatschke show that three very
different sets of the initial and freezeout temperature ($T_{\rm i}$,$T_{\rm f}$) ---
$(0.29, 0.14)$ GeV with $\tau_0=2$ fm, $(0.36, 0.15)$ GeV  
with $\tau_0=1$ fm, and $(0.43, 0.16)$ GeV with $\tau_0=0.5$ fm ---
provide almost identical differential elliptic flows in their viscous hydrodynamical  
calculation \cite{Luzum:2008cw}.
This suggests that better constraints on initial conditions
are indispensable to avoid incorrect conclusions from comparisons of hydrodynamical 
calculations with experimental data.

Other approaches have been also taken in generating initial conditions. 
Color glass condensate-inspired initial conditions are becoming increasingly
popular (Tabs.\ \ref{table:ideal-hydro} and \ref{table:viscous-hydro}).
They feature larger eccentricities of the initial energy profile 
than Glauber-based models, which has significant implications on elliptic 
flow \cite{Hirano:2004en}.
In these models additional dissipation during the early quark-gluon plasma 
stage is needed in order to achieve agreement with experiments 
\cite{Hirano:2004en, Hirano:2005xf}.
Others models include the string rope model \cite{Csernai:2009zz} and the
pQCD + saturation model \cite{Eskola:2005ue}. In the latter the initial 
time $\tau_0$ is given by the inverse of the saturation scale, 
which is very small, i.e., $\tau=$0.18 (0.10) fm at RHIC (LHC).

More recently there is a push to implement effects of 
event-by-event fluctuations in the initial conditions. 
In the NEXSPHERIO hydro model each event is created by the event generator
NeXus \cite{Hama:2004rr}.  First they found that the existence of fluctuation in initial 
conditions improves the behavior of the elliptic flow as a function of the
rapidity in hydrodynamical 
calculation \cite{Andrade:2008xh}. They showed that the two artificial bumps  
in the elliptic flow as a function of the 
rapidity \cite{Hirano:2002ds} disappear if they take into 
account of initial fluctuations. 

\begin{figure}
\begin{minipage}[h]{7.2cm}
\centerline{\includegraphics[width=7.2 cm]{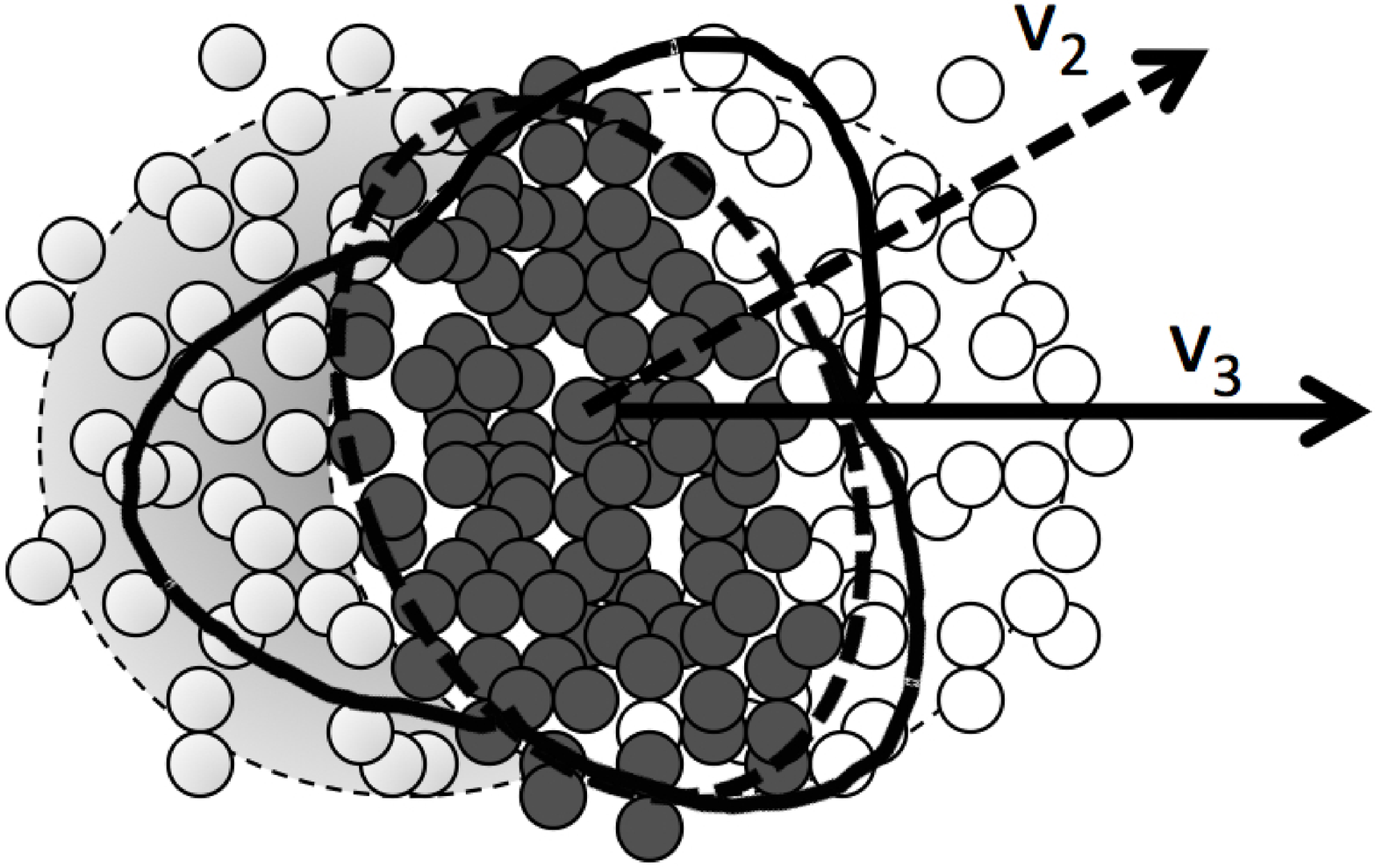}}
\caption{Elliptic and triangular flows from fluctuated initial state.}
\label{fig:v3}
\end{minipage}
\hspace{1cm}
\begin{minipage}[h]{7cm}
\centerline{\includegraphics[width=7.0 cm]{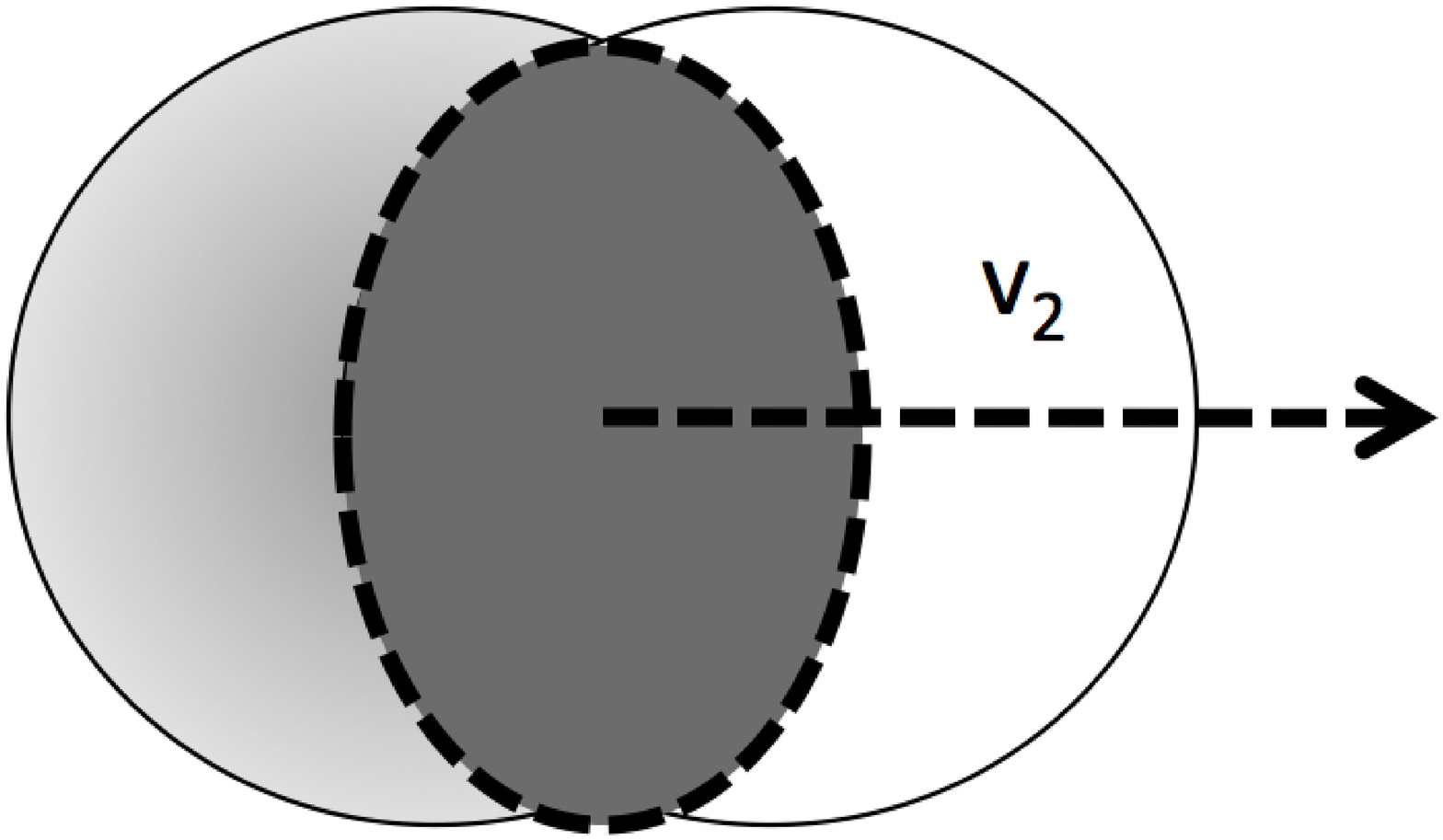}}
\caption{Elliptic flow from a smooth almond-shaped 
initial state.}
\label{fig:v2}
\end{minipage}
\end{figure}

An interesting observable, ``Mach-Cone-like structure", which is an angular
correlation with the leading jet particle
was reported at RHIC \cite{:2009qa,McCumber:2008id}.   
At first the origin of the structure
is considered as the remnant of the interactions between jets 
and medium ([18]--[26] in Ref. \cite{:2009qa}). 
Models succeeded in giving the qualitative interpretation of the structure, but  
they are not so successful in quantitatively describing the experimental data.   
For the first time in Ref. \cite{Takahashi:2009na} event-by-event
fluctuations in the initial state 
were considered as a possible origin of the structure.  
A breakthrough of clear understanding of the Mach-Cone-like structure was done by detailed 
experimental analyses of triangular flow and higher harmonics.  
Current understanding is that the Mach-Cone-like structure is dominated by the
triangular flow (See for example, Ref.\cite{ATLAS_vn}).  
The triangular flow and higher harmonics are the coefficients in 
the Fourier expansion of 
particle yields as a function of the azimuthal angle $\phi$,  
\begin{equation}
\frac{dN}{dyd\phi} \propto 1 + 2 v_1 \cos (\phi-\Theta_1) + 2 v_2 \cos 2(\phi-\Theta_2) 
+ 2 v_3 \cos 3(\phi-\Theta_3) + 2 v_4 \cos 4(\phi-\Theta_4) + \cdots ,  
\end{equation}
where  $v_1$, $v_2$, $v_3$, and $v_4$ are directed, elliptic, triangular, and 
quadrangular flows, respectively. $\Theta_i (i=1,2,\dots)$ are constants and
in principle independent from each other.
The origin of the triangular flow and higher harmonics is
fluctuation in initial conditions \cite{Han:2011iy} (Fig.\ \ref{fig:v3}). 
In particular, $v_3$ vanishes, if the system starts with a smooth almond-shaped
initial state \cite{Han:2011iy} (Fig.\ \ref{fig:v2}). 
Investigation of the relation between initial geometry and higher harmonics has been
also carried out  \cite{Gardim:2011xv}. 
For more quantitative analyses, however,  contributions of final state interactions 
should be evaluated. 

Here we make a comment on event-by-event fluctuations in the initial 
conditions in hydrodynamical models from the point of view of 
the numerical solution of the hydrodynamical equation. 
When the hydrodynamical simulation is performed with initial conditions
with the event-by-event fluctuation,
shock-wave capturing schemes should be used 
to describe the hydrodynamical expansion correctly.
Otherwise, the effect of the fluctuation is smeared.  It is
known that most of the numerical schemes used in the calculations
of the time evolution of the quark-gluon plasma do not satisfy
this requirement. In other words, they introduce numerical viscosity
to a non-negligible extent.
In the early stage of the RHIC operation, hydrodynamical models gave us 
certain evidence of the existence of strongly interacting QGP at RHIC and 
are eventually regarded as the most reliable dynamical models for high energy 
heavy ion collisions. 
However, recent high statistical experimental data impose more rigorous numerical  
treatment on the hydrodynamical models. 
We will discuss this issue in detail later in Sec. \ref{sec:NS}. 

\subsection{Experimental Data and Discussion}

Among various kinds of experimental observables, thermal photons are 
one of the most promising ones for the investigation of the initial
conditions of hydrodynamic models. PHENIX Collaboration reported that the excess from the 
superposition of pp collisions in its direct photon measurement at low transverse
momentum $P_T$. 
This excess follows the exponential distribution, which suggests that 
the thermal equilibrium is achieved \cite{:2008fqa}. 
They extract the inverse-slope parameter from the photon spectra in Au+Au 
$\sqrt{s}=200$ GeV, $T=221 \pm 19^{\rm stat} \pm 19^{\rm syst}$. 
This value is larger than the critical temperature of the phase 
transition ($\sim 160$ MeV), but it is smaller than the initial temperatures 
of hydrodynamic models at RHIC, $T\sim $ 300 -- 600 MeV. 
The value is interpreted as the average temperature in the whole hydrodynamic 
expansion. Contributions from the hadron phase are also important. 

It will be appropriate to point out here the importance of the hydro+micro
model as an instrument that relates the initial state condition to
the final state observables. Very often, for example, final state multiplicities
are directly predicted from the initial condition from the color glass
condensate picture. For example, when the first LHC multiplicity data came out 
\cite{Aamodt:2010cz}, deviation of the experimental data from
the prediction of the color glass 
condensate picture was found and was regarded as a serious problem. 
In fact, the system goes through several processes in which entropy
and multiplicity change, such as the thermalization of the quark-gluon plasma,
entropy production in the quark-gluon plasma, 
hadronization, and resonance production in the hadron phase and their decays.
It is thus to be emphasized that the understanding of the final state observables
requires the understanding of the whole period of the time evolution in heavy ion
collisions.
We will come back to this issue  in Sec.\ \ref{sec:freezeout}.

\section{Hydrodynamical expansion \label{sec:hydro}}
In the early stage of the RHIC operation, only hydrodynamical models
were able to explain 
the strong elliptic flow \cite{QM2001}, which was solid evidence of the creation of the strongly 
interacting QGP at RHIC. 
However, detailed analyses on experimental data eventually revealed the 
limitations of hydrodynamical models.
They have difficulty in explaining, for example, 
the elliptic flow at forward/backward rapidity, HBT results, 
the centrality dependence of elliptic flow, and so forth.
At the same time, viscosities became one of the most central topics in 
relativistic heavy ion collisions, while most of analyses had been carried out 
with ideal hydrodynamics.  
After the discovery of the strongly interacting QGP,
the main interest at RHIC shifted
to the understanding of 
detailed properties of the strongly interacting QGP. 
\subsection{Hydrodynamical equations}
The basis of hydrodynamical models is the energy and momentum conservation, 
\begin{equation}
\partial_{\mu}T^{\mu \nu}(x) = 0 \, , 
\label{eq-hydro}
\end{equation}
where $T^{\mu \nu}(x)$ is the energy momentum tensor. 
In the case of ideal relativistic fluid, the energy momentum tensor is given by 
\begin{equation}
T^{\mu \nu}(x)= [ \epsilon (x) + p(x) ] u^\mu (x) u^\nu  (x) - p(x) g^{\mu
    \nu} \, , 
\end{equation}
where $\epsilon (x)$, $p(x)$, and $u^\mu (x)$ are the energy density,
pressure, and four velocity, respectively. 
Eq. (\ref{eq-hydro}) is solved numerically simultaneously with the charge
conservation relation,  
\begin{equation}
  \partial_{\mu}j^\mu(x) = 0 \, .
  \label{eq-net_baryon}
\end{equation}

When one starts to include the effects of dissipation into relativistic hydrodynamics,
one is confronted with a rather complicated situation. 
One of the difficulties is that naive introduction of viscosities, 
first order theory (i.e., first order in gradients) suffers from
acausality. The heat conduction equation allows instantaneous propagation
of heat because of its parabolicity.
The acausality of first order hydrodynamics stems from the same reason.
In order to avoid this problem, second order terms in heat flow and 
viscosities have to be included in the expression for the entropy
\cite{Muller,Israel,IsSt1,IsSt2,GO1,GO2,Ott}, but the systematic treatment 
of these second order terms has not been established.
Although there is remarkable progress toward the construction of a 
fully consistent relativistic viscous hydrodynamical theory, there 
are still ongoing discussions about the formulation of 
the equations of motion and about the numerical procedures 
\cite{Teaney:2009qa}.

The basic tenet that has to be given up in dissipative hydrodynamics
is the assumption of a uniquely defined local rest frame. Away from 
equilibrium the vectors defining the flows of energy, momentum, and
conserved charges can be misaligned. We can still define a local rest
frame by just choosing a velocity $u^\mu(x)$ in the laboratory frame. Then
the energy-momentum tensor and the conserved charge current take more
general involved forms,
\begin{align}
    T^{\mu \nu}(x)& = [ \epsilon(x) + p(x)+\Pi(x) ] u^\mu(x) u^\nu(x)  - 
      [p(x)+\Pi(x)] g^{\mu \nu} + 2 W^{(\mu} u^{\nu)} + \pi^{\mu\nu} \, , \\
    j^\mu (x) &= n(x) u^\mu + V^\mu \, , 
\end{align}
where $V^\mu$ and $W^\mu$ are corrections to the flow of conserved
charge and energy that are orthogonal to $u^\mu$ and $T^{\mu\nu} u_\nu$,
respectively,
$\pi^{\mu\nu}$ (with the orthogonality conditions 
$u_\mu \pi^{\mu\nu}=\pi^{\mu\nu}u_\nu =0$) is the symmetric traceless shear 
stress tensor, and $\Pi$ is the bulk stress. $(\cdots )$
indicates the symmetrization with regard to the indices.
Usually $u^\mu$ is chosen to define one of the two standard frames: the Eckart 
frame where the velocity is given by the physical flow of net charge 
(then $V^\mu = 0$), or the Landau frame where the velocity is given by
the energy flow (then $W^\mu = 0$).  
We refer the reader to the article by Muronga and Rischke for further
discussions \cite{Muronga:2004sf}.

At first order the new structures are proportional to gradients
of the velocity field $u^\mu$, and only three proportionality constants appear:
the shear viscosity $\eta$, the bulk viscosity $\zeta$,
and the heat conductivity $\kappa$. 
With the usual
definitions the first order relations in the Landau frame are \cite{Muronga:2004sf}
\begin{align}
  \Pi &= -\zeta \nabla_\mu u^\mu \, ,   \\
  q^\mu &= - \kappa \frac{nT^2}{e+p} \nabla^\mu \frac{\mu}{T}\, , \\
  \pi^{\mu\nu} &= 2 \eta \nabla^{<\mu} u^{\nu>} \, ,
\end{align}
where $q^\mu = -(\epsilon+p)/n \, V^\mu$ is the heat flow,
$\nabla^\mu = (g^{\mu\nu} -u^\mu u^\nu) \partial_\nu$ is the covariant
derivative orthogonalized to the flow vector, $T$ and $\mu$ are the temperature
and chemical potential for the conserved charge, respectively.
Here only one conserved charge is assumed, but the extension to cases with
more than two conserved charges is straightforward.
$\langle \cdots \rangle$
refers to the symmetrization of indices with the trace subtracted. 
The entropy current $S^\mu$ receives additional contributions beyond the
equilibrium term $s u^\mu$ and one can show that all
three transport coefficients are positive, demanding that the entropy is strictly non-decreasing,
$\partial_\mu S^\mu \ge 0$.

At second order many more new parameters, related to relaxation phenomena,
appear.  Currently, most of viscous hydrodynamical calculations use the relativistic 
dissipative equations of motion that were derived phenomenologically 
by Israel and Stewart \cite{IsSt2} and variants of those, while some use the
method by \"Ottinger and Grmela \cite{GO1,GO2,Ott}. See for example Ref \ \cite{Dusling:2007gi}.
Recently, a second-order viscous hydrodynamics from AdS/CFT 
correspondence was derived \cite{Baier:2007ix}, as well as a set of 
generalized Israel-Stewart equations from kinetic theory via Grad's 14-momentum 
expansion, which have several new terms \cite{Betz:2009zz}. 
On the other hand, however, a stable first-order relativistic dissipative 
hydrodynamical scheme was also proposed on the basis of 
renormalization-group consideration\cite{Tsumura:2007ji,Tsumura:2007wu}.

In heavy ion physics, the shear viscosity, in particular its ratio with
the entropy density, $\eta/s$, has been attracted most of the attention among the
above three transport constants. 
Interesting seminal investigations on the effects of bulk viscosity have begun 
\cite{Fries:2008ts,Song:2009rh}, while heat conductivity still has not been 
investigated systematically in connection with RHIC data.
As for the second order parameters, there has been only little systematic
investigations, either.

Once equations of state are given, Eq.\ (\ref{eq-hydro}) is solved, the effect
of the phase transition being automatically taken into account, which is 
one of the advantages of the hydrodynamical model. We will discuss this feature in the 
next subsection. 
Recently the lattice(-inspired) equation of state which is connected to 
equation of state of resonance gas at low temperature is mostly used in 
hydrodynamical models.
As the fireball expands, the temperature and density inside become so small
that the assumption of the hydrodynamical picture becomes inapplicable any more.  
Thus, models with only a hydrodynamical component cannot 
describe the whole stages of the relativistic heavy ion collisions
from the thermalized quark-gluon plasma stage to the kinetic freezeout.
One way for such description is to connect a hydrodynamical simulation
to a hadron based event generator as discussed in Sec.\ \ref{sec:freezeout}. 

\subsection{Equation of State and Transport Coefficients \label{subsec:EoS}}
One of the advantages of hydrodynamical models over other phenomenological models is their
direct relation with the equation of state of QCD. Using the hydrodynamical 
models one can find directly the consequence of the phase transition
in experimental observables.
In hydrodynamical models, historically, the equation of state with a
first order phase transition based on the bag model widely has been used, 
because of its simplicity and lack of conclusive results on the equation of state of QCD.
In recent hydrodynamical calculations, lattice(-inspired) equation of state has been widely employed, because 
of the development of thermodynamical analyses based on the first principle calculation, 
lattice QCD simulation.
The equation of state of QCD for 2+1 flavors and also that including charm quark (2+1+1 flavors)  
by means of lattice simulations were reported by Borsanyi et al. \cite{Borsanyi:2010cj}.
HoTQCD collaboration investigates chiral and deconfinement aspects of the
phase transition with 2+1 flavors 
using several kinds of staggered fermions \cite{Bazavov:2011nk}.  
There was some difference between the two groups in the
critical temperature, the temperature dependence of 
the so-called interaction measure and so on, but gradually the difference is disappearing.
At the same time Wilson fermion simulation is also applied to the analysis of QCD
thermodynamical properties \cite{Umeda:2010ye}. 
Quantitative analyses of the QCD thermodynamics with the lattice simulation have just started. 
For conclusive results, improvement in actions, approach to continuum limit,
simulation at the real pion mass and so on need to be done. 

Simulations at finite chemical potential on the lattice suffer difficulty in execution of Monte Carlo 
simulations owing to the sign problem in the fermion determinant. 
Although there have been some developments recently such as the Taylor expansion method and
reweighting method \cite{FoKa02,FoKa,FoKaSz,FoKa2004,Ej,AlEj,AlEjHa,AlDo,GaGu,Whot-qcd,Nagata:2012pc},
the development in the study of finite chemical potential lattice simulations is
relatively slow compared to lattice QCD study at the vanishing chemical potential.  
Though a lot of attempts to circumvent the problem have been made, we still need a 
breakthrough to explore the QCD properties in the whole region on the $T$-$\mu$ plane.

In relativistic viscous hydrodynamical models, we need to input more information related to transport 
coefficients, shear and bulk viscosities, 
heat conductivity, and relaxation times, besides equations of state. 
The investigation of the transport coefficients of strongly interacting QGP
and hadron gas is one of
the most difficult problems in the field. 
There are several studies on transport coefficients  
with various approaches; AdS/CFT \cite{Baier:2007ix}, lattice QCD \cite{Nakamura:2004sy, Meyer:2007ic}, 
finite temperature perturbative QCD \cite{Arnold:2000dr, Arnold:2003zc,  Arnold:2006fz},
microscopic transport models \cite{Muronga:2003tb,Muroya:2004pu,Demir:2008tr,Xu:2007ns,Xu:2007jv},
and relativistic quantum Boltzmann approach \cite{Itakura:2007mx}. 
However, to reach a conclusive result on the transport coefficients, in addition to theoretical
calculations, it plays an important role to extract transport coefficients
from comparison of phenomenological model analyses and experimental data.

\subsection{Jet Energy Loss}
One of the most interesting features that are related to hydrodynamical expansion is the jet energy loss.  
A lot of interesting experimental results which suggest the existence of very large jet energy loss
are reported. To explain these results interaction between jets and medium need to be understood.
At the moment, there are perturbative QCD based 
approaches, the higher twist formalism, 
the AMY formalism, the GLV formalism, and ASW formalism,
and AdS/QCD approach for the jet energy loss (References are found, for example,
in Ref.\ \cite{Fries:2010ht}).

Despite the large amount of effort put into the development of
perturbative description of the hadron production in heavy ion collisions, there
are uncertainties remaining about the exact nature of jet-medium interactions
in the kinematic and temperature regimes relevant at RHIC. 
As a whole, the above four approaches describe RHIC data well, 
but they 
reach very different quantitative conclusions about the quenching strength,
or transport coefficient,  
$\hat q$. 
This does not come as a big surprise since the approaches differ in some 
of their basic assumptions, and there are large uncertainties in
modeling hard probes beyond the calculation of the energy loss rate for
a quark or gluon. 

Currently the big picture can be summarized as follows:
perturbative calculations under various assumptions are compatible with RHIC
data, but the constraints are insufficient to rule out any of the models. The
experimental constraints
are also insufficient to completely exclude non-perturbative mechanisms for jet quenching.
Calculations using the AdS/CFT correspondence to model strongly interacting
QCD \cite{Herzog:2006gh,Liu:2006ug,Gubser:2006bz} can describe the
same basic phenomenology. Most likely this challenge to perturbative QCD can only
be answered at LHC. The extrapolation of jet quenching to larger jet energies
is significantly different in strong coupling and perturbative scenarios 
\cite{Horowitz:2007su}. It is also possible to assume a small regime of
strong non-perturbative quenching around $T_{\rm c}$ together with the perturbative
quenching at higher temperatures. Such hybrid scenarios might be hard to
distinguish experimentally. One of such pictures was recently explored by Liao and Shuryak 
\cite{Liao:2008dk}. They found that a ``shell''-like quenching profile in
which quenching is enhanced around $T_{\rm c}$ can give better simultaneous 
fits to single hadron suppression and elliptic flow.
For more details, see, for example, Ref. \cite{Fries:2010ht}. 
\begin{table}[tb]
\begin{center}
\begin{minipage}{14cm}
\caption{pQCD-based energy loss models \cite{Fries:2010ht}: This table summarizes some of the key 
assumptions of the four perturbative calculations discussed in the text. The models
differ in the assumption on
the medium (thermalized, perturbative), kinematics,
scales ($E$ = energy of the parton, $k_T$ = transverse momentum of the
emitted gluon, $\mu$ = typical transverse momentum picked up from the medium,
$T$ = temperature, $\Lambda$ = typical momentum scale of the (non-thermalized)
medium, $x$ = typical momentum fraction of the emitted gluon), and the
treatment of the resummation. }
\label{tab:eloss}
\end{minipage}
\end{center}
\renewcommand{\arraystretch}{3}
\begin{center}
\begin{tabular}{|c|c|c|c|} 
\hline \hline
Model & Assumption about the medium and kinematics& Scales & Resummation \\ \hline\hline
GLV & \begin{minipage}[b]{6cm}static scattering centers (Yukawa), opacity
  expansion\end{minipage} & $E \gg k_T \sim \mu$,
$x \ll 1$ & Poisson
\\ \hline 
ASW & \begin{minipage}[b]{6cm}static scattering centers, multiple soft scattering (harmonic oscillator
approximation)\end{minipage}  & $E \gg k_T \sim \mu$, $x \ll 1$ & Poisson
\\ \hline
HT & \begin{minipage}[b]{6cm} observable matrix elements at scale $\Lambda$ (thermalized or non-thermalized medium) \end{minipage}  &
$E \gg k_T \gg \Lambda \sim \mu$ &  DGLAP
\\ \hline
AMY & \begin{minipage}[b]{6cm}perturbative, thermal, $g << 1$ (asymptotically large $T$)\end{minipage}  & $ E > T \gg
gT \sim \mu $ & Fokker-Planck
\\ \hline
\end{tabular}
\end{center}
\end{table}

%
\subsection{Hydrodynamical Expansion}
\begin{figure}
\hspace{-0.5cm}
\begin{minipage}[h]{7.5cm}
\centerline{\includegraphics[width=7.5 cm]{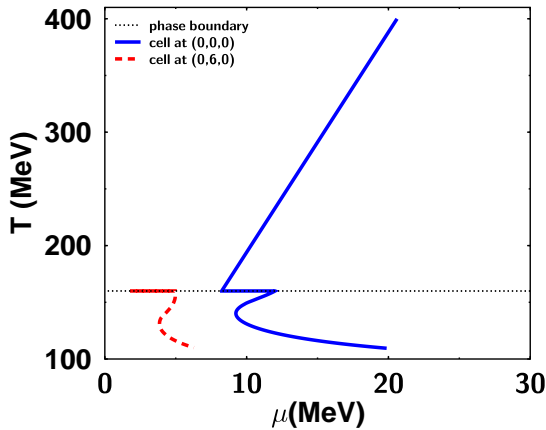}}
\end{minipage}
\begin{minipage}[h]{6cm}
\caption{Isentropic trajectories on the $T$|$\mu$ plane 
in the case of the 1st order phase transition. 
Solid (dashed) line stands for the time evolution of the
cell which is located at 
$(x, y, \eta) = (0, 0, 0)(= (0, 6, 0))$ at the initial time. 
The dotted line represents the phase boundary.
}
\label{fig:t-mu}
\end{minipage}
\end{figure}

Before going to the discussion of experimental data on
hydrodynamical expansion, in this subsection, we show the behavior of 
the temperature and chemical potential in hydrodynamical expansion. 
As an example, we depict the behavior of isentropic trajectories  
in the $T-\mu$ plane  
for Au+Au $\sqrt{s_{NN}}=200$ GeV central collisions 
in Fig.~\ref{fig:t-mu}. 
The dotted line stands for the phase boundary between the QGP and the hadronic phase
(Note that due to small baryochemical potentials, 
the phase boundary is an almost flat line at $T_{\rm c}=160$ MeV).  
In addition to the central cell, we also investigate the isentropic trajectory
of a cell close to the surface of the initially produced QGP.
Whereas the isentropic trajectory of the central cell located at 
$(0,0,0)$ starts in the QGP phase (solid line), 
the cell at the initial surface of the QGP (dashed line) only exhibits
an evolution from the mixed phase to the hadronic phase. 
Both trajectories are terminated at the freezeout temperature, 
$T_{\rm f}=110$ MeV.  

\subsection{Experimental Data and Discussion}
Calculated results shown in this subsection are based on Ref.\cite{Nonaka:2006yn}. 
To emphasize the importance of the final state interactions in the freezeout process
in understanding hadron observables in relativistic heavy ion collisions,
in this subsection we only show 
results of  pure hydrodynamical calculation. 
The calculation is performed with a (3+1) dimensional ideal hydrodynamical model with 
the Glauber type initial condition and a bag model equation of state. 
Effects of final state interactions will be discussed in Sec.\ \ref{sec:freezeout}. 
This calculation, thus, can be considered as a baseline for
recent more realistic hydrodynamical models.  

In the initial energy density distribution, the maximum value of the energy density is  
55 GeV/fm$^3$, which is relatively higher than in other hydrodynamical models, 
because in the pure hydrodynamical calculation we just use a single freezeout 
temperature and neglect both resonance decays and final state interactions. 
Usually parameters for the initial condition are set from comparison 
with experimental data of single particle distributions, rapidity distributions and
$P_T$ distributions in central collisions. Therefore hydrodynamical models 
have prediction power for other physical observables such as collective flow 
and the impact parameter dependence of various physical observables. 
We include the small baryon number density in these calculations. 
The starting time of hydrodynamical expansion is $\tau_0$ =0.6 fm. 
For the details of the parameters used in the calculations and the equation of state,
see Ref. \cite{Nonaka:2006yn}. 

\begin{figure}
\begin{minipage}[h]{7cm}
\centerline{\includegraphics[width=7 cm]{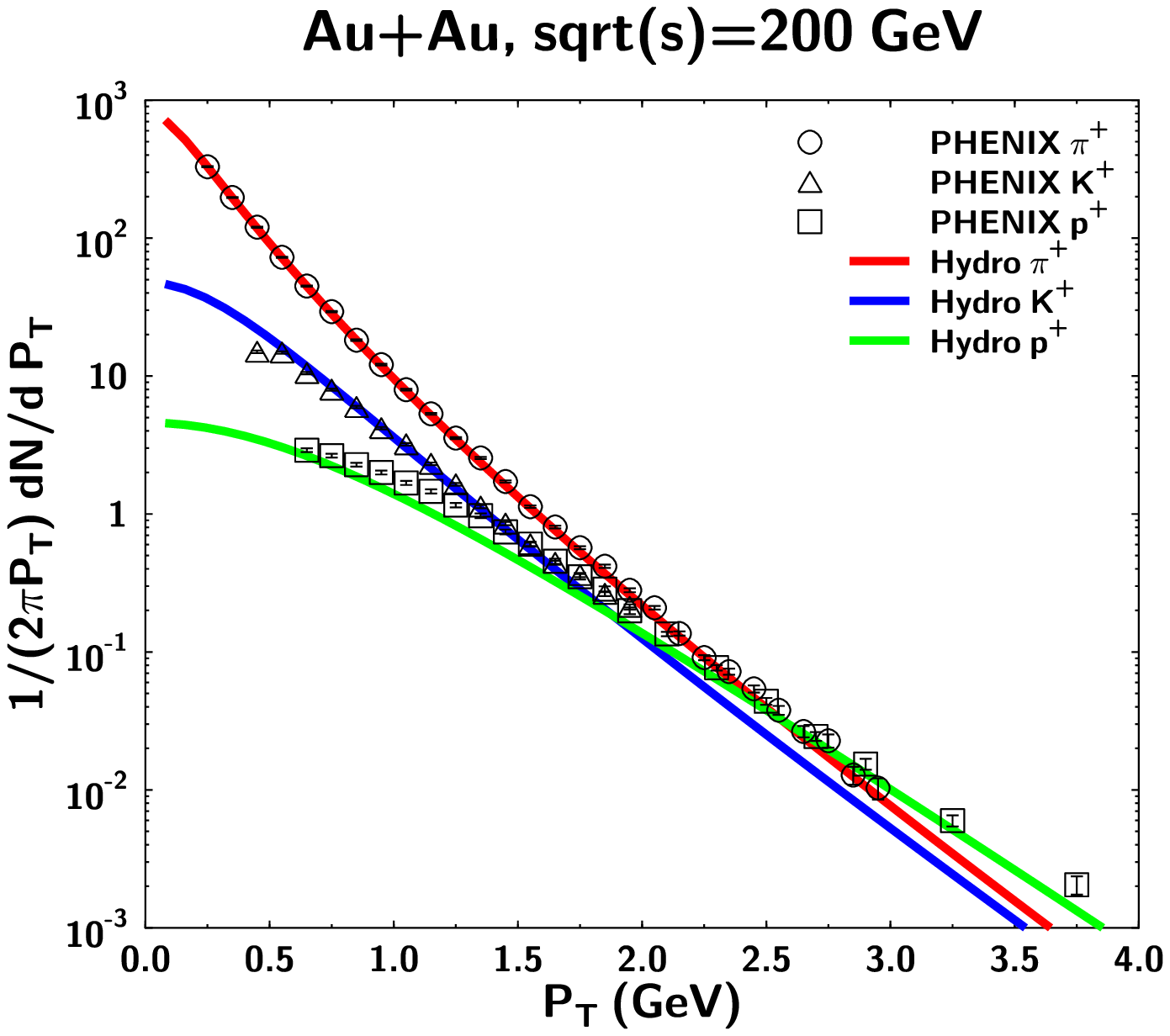}}
\caption{ $P_T$ spectra for $\pi^+$, $K^+$, and $p$ in central collisions 
with PHENIX data\cite{PHENIX_pt}. 
For the proton we rescale our result using the ratio at the chemical freezeout temperature.
See the text.}
\label{fig:pt-h}
\end{minipage}
\hspace{1cm}
\begin{minipage}[h]{7cm}
\centerline{\includegraphics[width=7.3 cm]{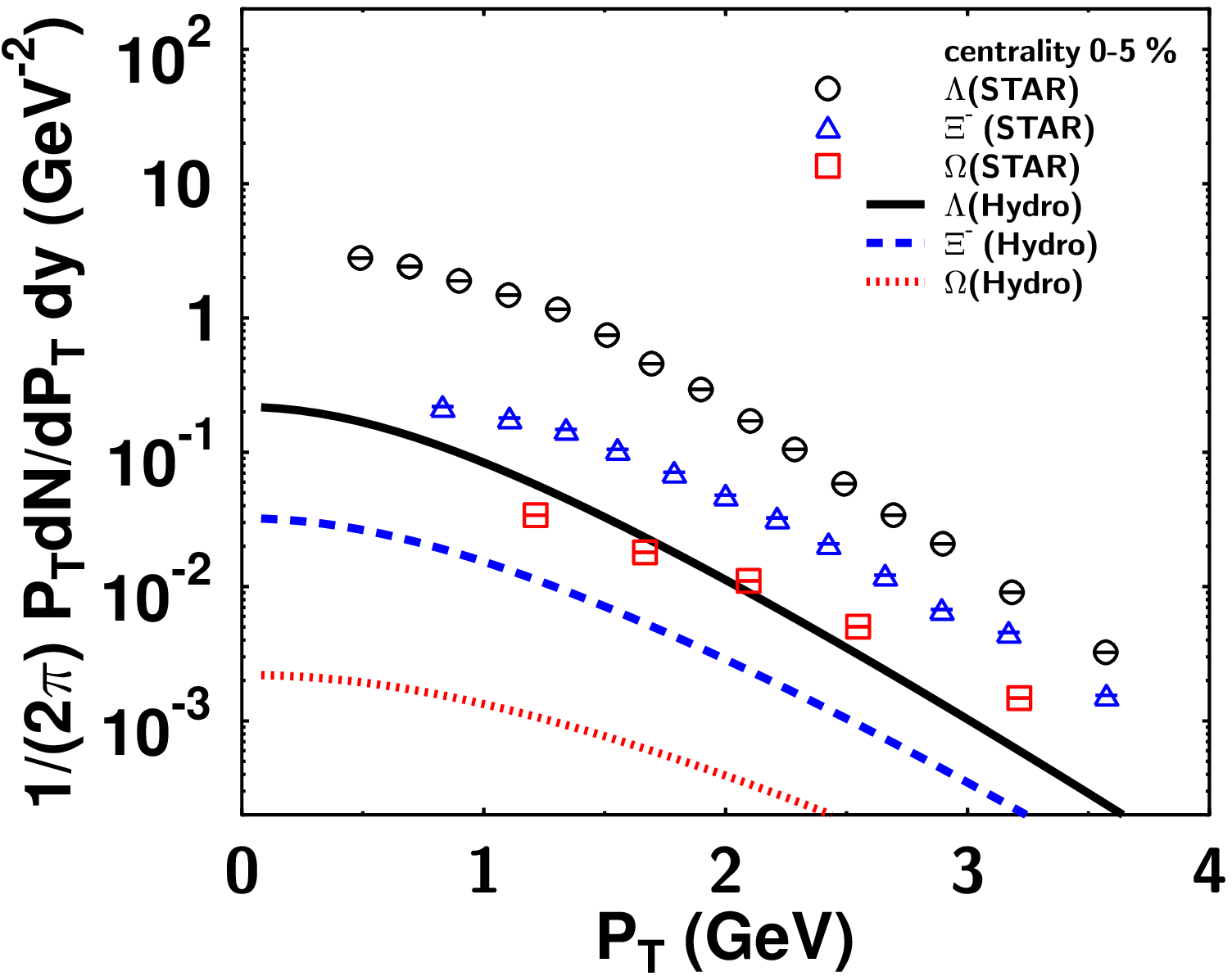}}
\caption{$P_T$ spectra for multistrange baryons in central 
collisions with STAR data \cite{STAR_pt_strange}. 
In this pure hydro calculation no additional procedure for normalization 
has been done.}
\label{fig:pt-s-h}
\end{minipage}
\end{figure}

First we show two examples which clearly show the 
limitation of pure hydrodynamical models. 
One is the $P_T$ spectra of $p$ and multistrange particles and 
the other is the elliptic flow as a function of the rapidity.  
Figure \ref{fig:pt-h} shows the 
$P_T$ spectra of $\pi$, $K$, and $p$ in Au + Au at $\sqrt{s_{\rm NN}}=200$ GeV 
for central collisions. 
Our calculation succeeds in reproducing the $\pi$ spectra measured    
by PHENIX \cite{PHENIX_pt} up to $P_T \sim 2$ GeV. 
However, due to the model assumption of chemical 
equilibrium down to the (low) kinetic freezeout temperature, 
we fail to obtain the correct normalization and hadron number ratios, 
even though the shape of the $P_T$ spectra of $p$ and multistrange 
baryons (shown in Fig.~\ref{fig:pt-s-h}) is close to that of the
experimental data. 
In order to obtain the proper normalization for the $p$
spectra and hadron number ratios,
we adopt a procedure outlined in Ref.~\cite{hadron-ratio};
it renormalizes the proton $P_T$ spectra using the $p$ to $\pi$ ratio 
at the critical temperature.
It is straightforward to extend this procedure to hyperons
and multi-strange baryons as well, even though we choose to show
the raw, unrenormalized, result for the multi-strange baryons
in Fig. \ref{fig:pt-s-h} to elucidate the situation prior to
the renormalization.

The need for renormalizing the $p$ spectra 
suggests that the assumption of a persistent chemical 
equilibrium throughout the hadron phase until kinetic freezeout  
is not realistic and that an improved treatment of the freezeout process 
is required. 
One method to deal with the separation of chemical and
thermal freezeout is the {\em partial chemical equilibrium model (PCE)}
\cite{Hirano:2002ds,Te02,Kolb:2002ve}:
below the chemical freezeout temperature $T_{\rm ch}$ one
introduces a chemical potential for each hadron whose
yield is supposed to be frozen out at that temperature.
While the PCE approach can account for the proper normalization
of the spectra, it fails to reproduce the
transverse momentum spectra and mass dependence of the elliptic 
flow \cite{PHENIX_white}.  
In Sec.~\ref{sec:freezeout}, 
we shall utilize our hybrid hydro+micro model 
to decouple the chemical freezeout from the kinetic freezeout. In these
hybrid approaches
\cite{BaDu00,TeLaSh01,Hirano:2005xf}, 
the freezeouts occur
sequentially as a result of the microscopic evolution. 
Flavor degrees of freedom are treated explicitly through the
hadronic cross sections in the microscopic transport, which leads
to the proper normalization of all hadron spectra.

\begin{figure}
\begin{minipage}[h]{7cm}
\centerline{\includegraphics[width=7 cm]{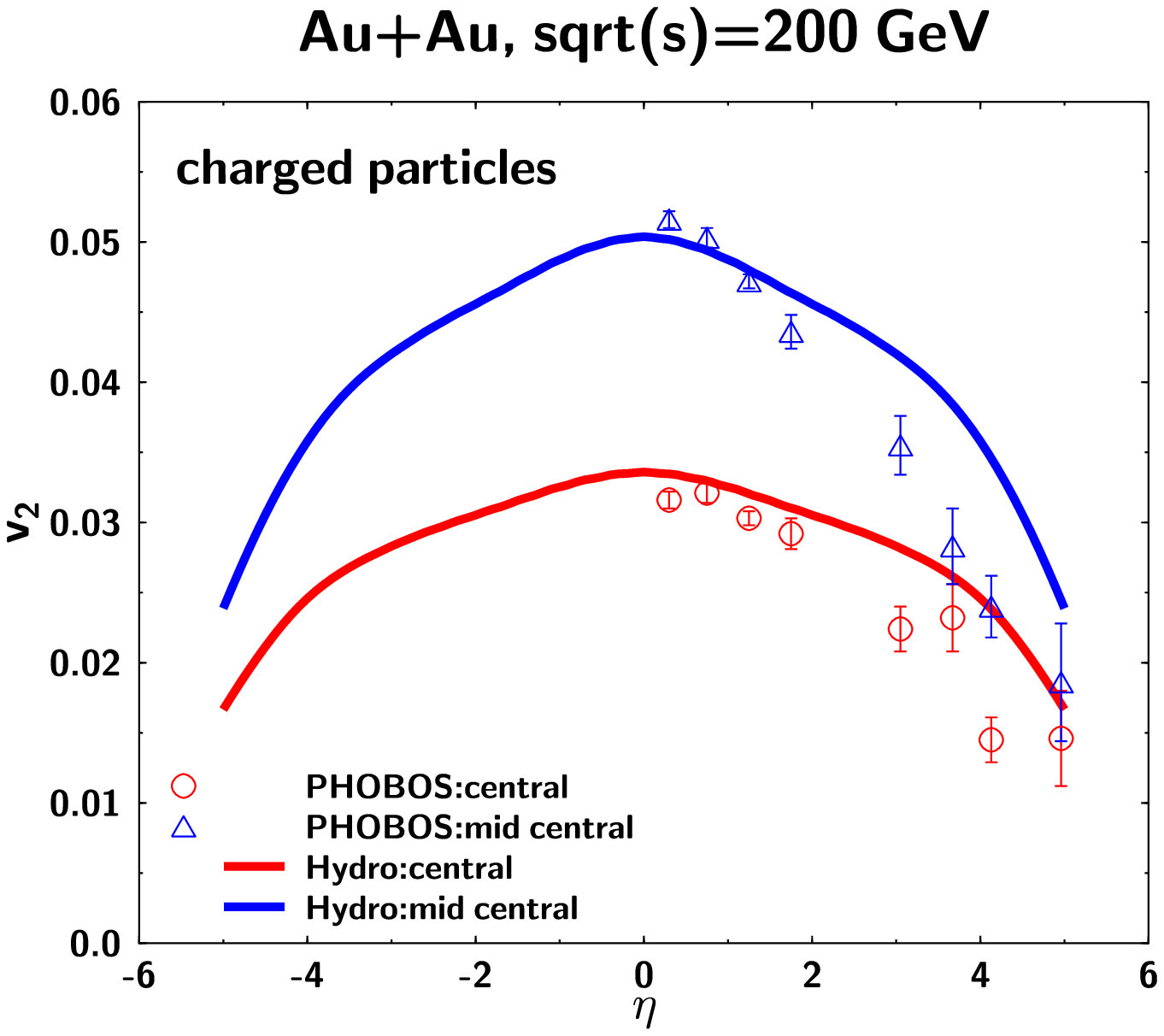}}
\caption{Elliptic flow as a function of $\eta$ with PHOBOS 
experimental data \cite{PHOBOS_v2_eta} for central (3-15 \%) and mid 
central (15-25\%) collisions. 
Impact parameters are set to 4.5 (central) and 6.3 (mid central) fm, 
respectively. }  
\label{fig:v2-eta-h}
\end{minipage}
\hspace{1cm}
\begin{minipage}[h]{7cm}
\centerline{\includegraphics[width=7.3 cm]{./fig/bass_RAA_centrality.eps}}
\caption{$R_{AA}$ as a function of $P_T$ for central (top) and mid-central 
(bottom) collisions calculated from ASW, Higher Twist, and AMY energy loss models.
The single parameter in each model has been fitted to describe the data by PHENIX\cite{PHENIX_RAA}. }
\label{fig:RAAcomparison}
\end{minipage}
\end{figure}

Figure \ref{fig:v2-eta-h} shows the elliptic flow as a function of the pseudo-rapidity
$\eta$ in central (3-15 \%) and mid central collisions (15-25 \%). 
In both cases our hydrodynamical  
model calculations overestimate the elliptic flow at forward and backward pseudo-rapidities, 
similarly to the results shown in
Ref.~\cite{Hirano:2002ds}.    
At large forward and backward rapidities the assumptions  
of perfect hydrodynamical models such as local equilibrium, 
vanishing mean free path, and negligible viscosity effect,
are apparently no longer valid. 
The difference between experimental 
data and calculated results at forward and backward rapidities
increases with the impact parameter, implying
decrease of the volume in which
the hydrodynamical limit is achieved.

Next we move on to the topic of jet energy loss in expanding hydrodynamical medium
as one of interesting applications of hydrodynamical model. 
Figure \ref{fig:RAAcomparison} shows results from a comparative study by 
Bass et al.\ at RHIC\ \cite{Bass:2008rv}. Jets are propagated through a medium described
by the same
hydrodynamics, using three different schemes for energy loss: ASW, Higher Twist (HT), and
AMY. 
In Fig.\ \ref{fig:RAAcomparison} $R_{AA}$ as a function of $P_T$
for two different centrality bins is shown. It shows that 
the $P_T$-dependence and the centrality dependence of $R_{AA}$ are
described well by all three models. Each model has one free
parameter that has been adjusted: the strong coupling $\alpha_S$ in AMY,
$\hat{q}_0$ for the overall fit parameter in HT, $K$ ($\hat{q} =K \cdot 2 \cdot \epsilon^{3/4} $) 
in ASW. These parameters are fixed from the comparison with $R_{AA}$ data in 
central collisions (the top figure in Fig.\ref{fig:RAAcomparison}).

This study confirms the remarkably large $\hat q$ in the ASW model
compared to that in the HT approach. For the case where the quenching strength
scales with $\epsilon^{3/4}$ the initial values $\hat q_0$\footnote{Note
that the meaning of $\hat q_0$ in Eqs. (\ref{eq:qhat-e}) and (\ref{eq:qhat-t})
is different from the overall parameter $\hat q_0$ in HT. Both $\hat q_0$
are commonly used in the literature.} 
found for the quark at the 
center of the fireball in a central collision are \cite{Bass:2008rv}
\begin{equation}
  \hat q_0 = 18.5\>  \text{GeV}^2/\text{fm for ASW}, \qquad
  \hat q_0 = 4.5 \> \text{GeV}^2/\text{fm for HT}
\label{eq:qhat-e}
\end{equation}
and for the case where the quenching strength scales like the temperature $T$
they are
 \begin{equation}
  \hat q_0 = 10\>  \text{GeV}^2/\text{fm for ASW}, \hspace{2mm}
  \hat q_0 = 2.3\>  \text{GeV}^2/\text{fm for HT}, \hspace{2mm} 
  \hat q_0 = 4.1\>  \text{GeV}^2/\text{fm for AMY} .
\label{eq:qhat-t}
\end{equation}
Note that the jet propagation in AMY model is calculated self-consistently as a function
of the local temperature so that there is no difference between the two cases.

This comparison is unique and very valuable in the respect that the same initial
hard cross sections and the same maps for the fireball, from the same
(3+1)-dimensional ideal hydrodynamics were used. Any differences in the extracted values of 
$\hat q$ are solely due to differences in models for the jet energy loss.
One of the conclusions is that our current knowledge
applied to $R_{AA}$ leaves rather large uncertainty in the determination
of $\hat q$.

Here we make a short comment on the jet energy loss at LHC. 
ALICE collaboration reported the nuclear modification factors $R_{AA}$ as a function of $P_T$ of 
charged particles in central and peripheral Pb-Pb collisions at $\sqrt{s_{NN}}=2.76$ TeV 
\cite{Aamodt:2010jd}. 
They found that in central collision 
$R_{AA}$ is more suppressed up to 
$P_T=7$ GeV compared to the PHENIX and 
STAR experiments at RHIC, which suggests that a much denser medium is produced 
and stronger parton energy loss occurs at LHC. 
$R_{AA}$ decreases with $P_T$ for $2< P_T< 7$ GeV and it takes the 
minimum around $P_T = 7$ GeV and 
interestingly, increases with $P_T$ for $P_T > 7$ GeV.  
The increase of $R_{AA} $ at high $P_T$ was observed, for the first time, at LHC.
It was not seen clearly at RHIC.

%
%

\section{Numerical Schemes for Solving Hydrodynamical Equations \label{sec:NS}}
In this section, we first explain the current status of hydrodynamical models 
from the point of view of numerical schemes for relativistic hydrodynamical 
equations. 
As we showed in Tabs.~\ref{table:ideal-hydro} and \ref{table:viscous-hydro},   
hydrodynamical models are categorized into 
ideal  and viscous hydrodynamical models. 
In addition, the difference of each hydrodynamical model is found in 
the space-time dimension of simulation, 
initial conditions, equations of state, and prescriptions for freezeout process.  
In current understanding, the most realistic hydrodynamical model should have 
the following features: viscosity effects, (3+1) dimensional space-time expansion, 
event-by-event fluctuated initial conditions, lattice QCD inspired equations of state, 
and freezeout process which is described by hadron based cascade models. 
On these issues ideal hydrodynamical models have been studied
more deeply and its status is considered to be more mature than viscous 
hydrodynamical models. 
The investigation with viscous hydrodynamical models has just started. 

In addition to the above issues, an important ingredient in hydrodynamical
models should be taken into account seriously. It is what numerical scheme
is adopted for solving the relativistic hydrodynamical equation. 
Up to now, only a little attention has been paid to this point.
As long as we analyze multiplicities and collective flow using smoothed 
initial conditions, which numerical scheme to choose is not so important. 
However, when we start to investigate viscosity effects and event-by-event fluctuations,
we need to choose suitable numerical schemes carefully. 

The relativistic hydrodynamical equation is a non-linear system equation, whose 
analytical solution is usually difficult to find.
However, from the one dimensional wave equation $\partial_t u + c\partial_x u=0$,
which is much simpler than the actual relativistic hydrodynamical equation,
we can explore a suitable numerical scheme by comparing with the analytical solution. 
The naive differential scheme such as the FTCS (Forward-Time Central-Space) scheme 
($u^{n+1}_j=u^n_j - \frac{1}{2} \nu (u^n_{j+1}-u^n_{j-1}$)
\footnote{The upper index ($n$) stands 
for the time step and the lower index ($j$) stands for the spacial position.}, 
with $\nu=c\frac{\Delta t}{\Delta x}$) 
causes an unphysical oscillation, which continues to grow after several time steps. 
In order to stabilize the unphysical oscillation in the FTCS scheme, 
one can use, for example, the Lax-Friedrich scheme. 
In this scheme $u^n_j $ on the right hand side of the FTCS scheme is replaced 
by the averaged value $\frac{1}{2}(u^n_{j-1}+ u^n_{j+1})$. 
This scheme is stable if the Courant-Friedrichs-Lewy condition (CFL condition)  ($|\nu| < 1$) is satisfied, but it is known that 
it suffers huge numerical dissipation. In other words, the average manipulation introduces the 
artificial viscosity. 
One of the improved versions of Lax-Friedrich scheme is the Lax-Wendroff scheme, 
which has the second order accuracy in time and space, 
\begin{eqnarray}
u^{n+1/2}_{j+1/2} & = & \frac{u^n_{j+1}+ u^n_j}{2} - \frac{1}{2}c\frac{\Delta t}{\Delta x} 
(u^n_{j+1}-u^n_j), 
\nonumber \\
u^{n+1}_j & = & u^n_j - c\frac{\Delta t}{\Delta x} 
(u^{n+1/2}_{j+1/2}-u^{n-1/2}_{j-1/2}).  \nonumber 
\end{eqnarray}
This scheme is stable, but the unphysical oscillation occurs at discontinuity.
In order to avoid this problem introduction of artificial viscosity or flux limitation 
is needed.
In conjunction with this fact we just cite Godunov's theorem: 
no second-order or higher order explicit monotonous scheme exists. 
Systematic discussion on numerical schemes is out of scope of this article. 
For details, please see, for example,  Ref. \cite{Toro}.  
From these lessons for numerical schemes for the wave equation, we can 
deduce those for the relativistic hydrodynamical equation, though  
actual numerical tests are indispensable.
i) First order accuracy scheme: For stability at discontinuity 
some average manipulation is needed, but this introduces
large dissipation. ii) Second order accuracy: To remove numerical oscillation 
at discontinuity, one needs an artificial viscosity or a flux limiter.  
The former suggests that a simple central difference scheme
of the first order accuracy
might have a large artificial viscosity, which is crucial for the study
of the viscosities of the matter created in relativistic heavy ion collisions.  

\begin{figure}
\hspace{-0.5cm}
\begin{minipage}[h]{7cm}
\vspace{-0.4cm}
\centerline{\includegraphics[width=7 cm]{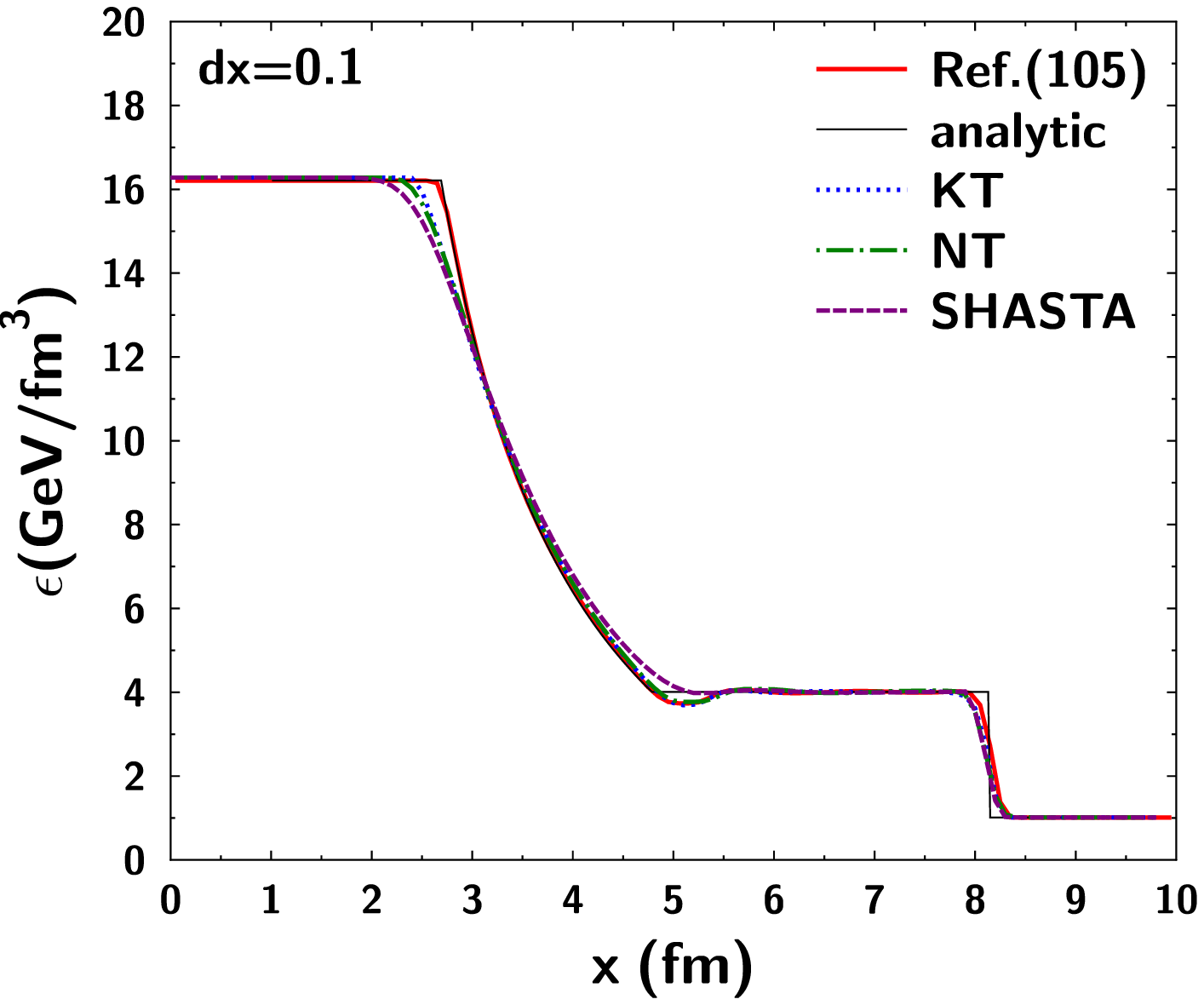}}
\caption{Energy density distributions as a function of $x$ from numerical 
results (SHASTA (short-dashed line), KT (dotted line), NT (long-dashed line), 
and Ref.\cite{AkNoTaIn} (red solid line)) together with the analytic Riemann solution. 
}
\label{fig:ene_ns}
\end{minipage}
\hspace{0.1cm}
\begin{minipage}[h]{7cm}
\vspace{-0.4cm}
\centerline{\includegraphics[width=7.0cm]{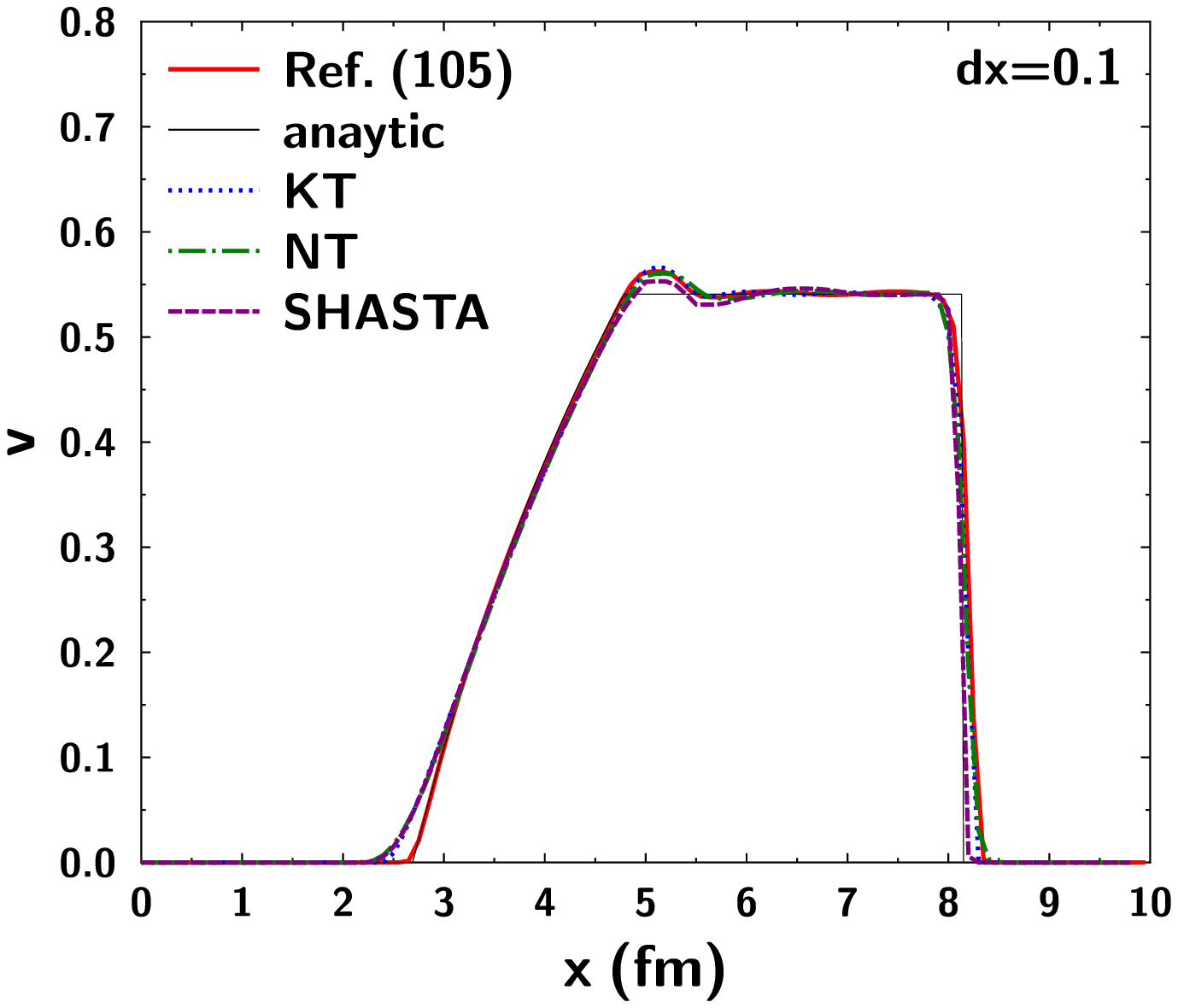}}
\caption{Velocity distributions as a function of $x$ from numerical 
results (SHASTA (short-dashed line), KT (dotted line), NT (long-dashed line), 
and Ref.\cite{AkNoTaIn} (red solid line)) together with the analytic Riemann solution. 
}
\label{fig:vx_ns}
\end{minipage}
\end{figure}

For event-by-event fluctuated initial conditions, shock-wave capturing 
schemes play an important role in dealing with discontinuity in the initial conditions.
A lot of shock-wave capturing schemes have been proposed and developed. 
On the other hand, in relativistic heavy ion collisions, SHASTA, rHLLE, and 
KT algorithms are mainly used. 
In particular, SHASTA algorithm, which is widely used in the study of relativistic 
heavy ion collisions, is known as the first version of Flux Corrected Transport (FCT) 
algorithm \cite{BoBo}. In this algorithm, first a low-order solution is calculated. 
It incorporates large numerical diffusion effect. In the second step, as much 
diffusion as possible is removed from the low-order solution. The amount of 
antidiffusion fluxes is determined with the mask coefficient $A_{\rm ad}$. 
This default value is $A_{\rm ad}=1$, which can be set to lower values to 
reduce the amount of antidiffusion.  
In Ref. \cite{Molnar:2009tx} an interesting demonstration of 
the interplay between the numerical viscosity and physical viscosity
was shown.
A comparison was made with numerical solutions of the Riemann 
problem, one with the standard mask coefficient $A_{\rm ad}=1.0 $ SHASTA 
with a small physical viscosity $\eta/s=0.01$ and the other
with a reduced mask coefficient 
$A_{\rm ad}=0.8$ SHASTA with vanishing viscosity.   
It was found that the difference of two numerical calculations is very small, 
which implies that it might be difficult to distinguish between the physical viscosity and 
the artificial dissipation. 
For quantitative investigation of the viscosities in the quark matter,
one need to estimate the amount of artificial viscosity existing in
numerical schemes carefully or choose numerical schemes known to have
less artificial viscosity. 
Furthermore in Ref. \cite{Molnar:2009tx},
a comparison among the different shock-wave capturing 
schemes, SHASTA, KT, and NT schemes was made and found
that all the algorithms reproduce the analytic solution with nearly the same 
accuracy and numerical artifact (Figs. \ref{fig:ene_ns} and \ref{fig:vx_ns}). 

Recently, Ref.~\cite{AkNoTaIn} proposed
a fast numerical scheme for causal relativistic hydrodynamics
with dissipation for analyses of relativistic high energy collisions,
which is based on Ref.~\cite{Takamoto:2011wi}.
In the numerical scheme,
Israel-Stewart theory was adopted and the Israel-Stewart 
equations were decomposed into the inviscid part and the dissipative part. 
For the inviscid part a relativistic Riemann solver 
\cite{MaMu94,Ma,AlMlMaMu,PoMaMu,FoMuSuTo,ZaBu,MaMu2003,MiBo,MiPlBo} 
is used and for the dissipative part 
the Piecewise Exact Solution method\cite{PES} is employed in order
to achieve less
artificial dissipation and less computational time. 
Riemann solvers are numerical fluid dynamical solvers
which are based on the concept of 
the Riemann problem. Several kinds of Riemann solvers have been 
proposed: Godunov method, Roe scheme, HLLE and HLLC solvers, and so on. 
Each solver has advantages and disadvantages in numerical cost, calculational 
accuracy, and coding complexity \cite{Toro, Numerical_HD}. 
In order to obtain the correct physical viscosity,
for the inviscid part as well as the dissipative part, we need to choose 
numerical schemes which suffer as little artificial dissipative effect 
as possible. In Ref.~\cite{AkNoTaIn}, for the inviscid part they use the 
relativistic Godunov method which is based on the exact solution of
the Riemann problem.
Figures \ref{fig:ene_ns} and \ref{fig:vx_ns} show numerical solutions of  
the relativistic Riemann problem for ideal fluid on a grid with $N_x$=100 cells 
with $\Delta x$=0.1 fm after $N_t=100$ time steps at 
$t=4$ fm together with the analytic solution. 
The initial conditions are set as the same as those in Ref.~\cite{Molnar:2009tx}. 
On the left the initial temperature is $T_0=0.4 $ GeV and on the right $T_0=0.2$ GeV,
and the energy density $\epsilon$
is given by $\epsilon=\frac{3g}{\pi^2}T^4$ $(g=16)$. 
The numerical results for SHASTA, KT, and NT schemes are taken 
from Ref.~\cite{Molnar:2009tx}. 
SHASTA, KT, and NT schemes reproduce the analytic Riemann solution with nearly 
the same accuracy and numerical artifacts. On the other hand, we can see that 
the red line\cite{AkNoTaIn} is closer to the analytical solution
around $x=3$ fm, which implies that 
this scheme suffers less artificial dissipative effect
and is more suitable for physical viscosity analyses. 
A comparison of rHLLE and SHASTA was done in Ref.\cite{Schneider:1993gd}. 
It was found that rHLLE has almost the same artificial viscosity as 
SHASTA.

\section{Hadronization \label{sec:hadronization}}
In hydrodynamical models the hadronization process from the QGP phase to the hadron phase
is naturally encoded in the equation of state. At RHIC hydrodynamical calculations
do well for explaining a lot of experimental data, the $P_T$  
spectra, and the elliptic flow up to  $P_T \sim$ 2 GeV.  
However, above $P_T \sim 2 $ GeV the results from hydrodynamical models
start to show deviation from experimental data, which suggests that other
mechanisms become dominant at higher $P_T$ instead of the hydrodynamical feature.  
In the intermediate $P_T$ ( $2 \leq P_T \leq 6$ GeV) the recombination mechanism is dominant and at high $P_T$  
( $\geq 6$  GeV) the fragmentation mechanism plays an important role. 
These transverse momentum regions where the hydrodynamical picture, recombination model, or fragmentation mechanism 
works well depend on collision energy. For example, at LHC the hydrodynamical description
may work up to $P_T\sim 4 - 5 $ GeV \cite{Eskola:2005ue}. 
There have been attempts to use thermodynamical properties which are described
with a hydrodynamical model as inputs 
of the recombination model and fragmentation mechanism\cite{Lee:2008bg,Fries:2011dj}. 
It may become possible to 
construct more realistic dynamical models
to understand the physics of relativistic heavy ion collisions  
by employing the recombination model and fragmentation process for hadronization
in a hydrodynamical model.  
In the next subsection, we give a brief description of the recombination model.

\subsection{Recombination Model}
Quark recombination or coalescence is the best candidate 
for the physical mechanism
to explain a large 
amount of experimental data at the intermediate $P_T$ at RHIC. 
Now not only at LHC but also at relatively low collisions energy experiments,  
analyses with the recombination picture are ongoing \cite{Hwa:2011bw}.   
First the quark number scaling law was found in the behavior of elliptic flow as a function of 
the transverse momentum, which is considered as a signature of the quark recombination. 
For a quark phase with the elliptic flow $v_2^q(P_T)$ at the time of the hadronization simple
instantaneous recombination models predict
\begin{equation}
  v_2^H(P_T) = n v_2^q\left(\frac{P_T}{n}\right) \, ,
\end{equation}
where $n$ is the number of the valence quarks. This scaling law describes the 
key feature of experimental data at the intermediate $P_T$ notably well.

Generally, recombination models assume
a universal phase space distribution of quarks at hadronization. Quarks turn
into baryons, $qqq \to B$, and mesons, $q\bar q\to M$. These processes
are described either by 
using instantaneous projections of quark states onto hadron states  
\cite{Fries:2003vb,Fries:2003kq,Greco:2003xt,Greco:2003mm,
Muller:2005pv}, or dynamical coalescence processes with finite width hadrons
governed by rate equations \cite{Ravagli:2007xx}.
Note that usually only the valence quarks of the hadron are taken into 
account although generalization has been attempted \cite{Muller:2005pv}.

The original instantaneous projection models explicitly preserve only three
components of the energy-momentum four-vector in the $2\to1$ and
$3\to 1$ processes. The yield of mesons can be 
expressed through the convolution of the Wigner function $W_{ab}$ for a pair of partons,
$a$ and $b$ and the Wigner function $\Phi_M$ encoding the meson wave function
\begin{equation}
  \frac{dN_M}{d^3 P} = \int \frac{d^3 R}{(2\pi)^3} \sum_{ab} \int
  \frac{d^3q d^3r}{(2\pi)^3} W_{ab}\left(\mathbf{R}+\frac{\mathbf{r}}{2},
  \frac{\mathbf{P}}{2}+\mathbf{q} ; \mathbf{R}-\frac{\mathbf{r}}{2},
  \frac{\mathbf{P}}{2}-\mathbf{q} \right)
  \Phi_M(\mathbf{r},\mathbf{q}) \, .
\end{equation}
The quark Wigner functions are usually approximated by the classical phase space
distributions. Hadron spectra at intermediate $P_T$ are described well by
considering the factorization of $W_{ab}$ into the thermal quark distributions
\cite{Fries:2003kq},
\begin{equation}
  W_{ab}\left(\mathbf{r}_1,\mathbf{p}_1;\mathbf{r}_2,\mathbf{p}_2\right)
  = f_a\left(\mathbf{r}_1,\mathbf{p}_1\right) f_b\left(\mathbf{r}_2,\mathbf{p}_2\right)
  \, . 
\end{equation}
Correlations between quarks can be introduced to model correlations found 
between hadrons \cite{Fries:2004hd} without interfering with the excellent 
description of the spectra and hadron ratios.
 
The strength of the quark recombination picture is in its predictive power, which
originates from its explaining all measured hadron spectra at the intermediate $P_T$
basically with one parameter for the quark distribution function at hadronization.
It has been shown that
at low momenta resonance recombination is compatible with hydrodynamics
and kinetic equilibrium \cite{He:2010vw}, but on the other hand, because
thermalized states do not retain memories of previous time, all
phenomena with long time scale in the equilibrated region should be expected to be described
by hydrodynamics. They include the quark number and kinetic energy scaling observed at RHIC 
at low momenta \cite{He:2010vw}.
The possibility of including quark recombination explicitly into
hydrodynamical model has been studied in Refs. \cite{Lee:2008bg,Fries:2011dj}.

\subsection{Experimental Data and Discussion}
Using the recombination model, we can explore the production of quark soup,
which is a consequence of the QCD phase transition. 
After the success of the recombination model at RHIC, the scaling property of 
the elliptic flow has been tested in a wide range of
collision energy to investigate the properties of 
the strongly interacting QGP. 
The elliptic flows of $\pi^{\pm}, p$, and $\Lambda$ were measured  
in Pb+Pb collisions at $\sqrt{s_{NN}}=17$ GeV at SPS by NA49 \cite{Alt:2006ye}.  
They found that the quark number scaling in the elliptic flows of these particles   
holds in the $P_T$ range covered by the data (up to $P_T/n \approx 1$ GeV. Fig. 6 in Ref. 
\cite{Alt:2006ye}). 
Since the $P_T$ is limited to rather low values, however, this
should not be viewed as a conclusive test for the quark recombination 
mechanism at SPS.

At LHC the analyses of the quark number scaling in the elliptic and triangular flows of 
identified particles have just started. 
In Ref.\cite{ALICE_v2}, ALICE shows
the elliptic and triangular flows per 
constituent quark of $\pi^{\pm}, K^{\pm}$, and $\bar{p}$ as a function of the
transverse kinetic energy ($KE_T$)
per quark for more central (10 -20 \%) and more peripheral (40-50 \%) 
Pb+Pb collisions at $\sqrt{s_{NN}}=2.76$ TeV. 
They report that within errors the flows of $\pi$ and $K$ follow the scaling, 
while the flow of $\bar{p}$ deviates in the more central and more peripheral 
events. In the triangular flow they find the same feature as the elliptic flow, 
i.e., the triangular flow of $\bar{p}$ shows deviations from the $KE_T$ scaling. 
However, this investigation is done in a relatively low transverse 
kinetic energy region ($KE_T/n < 1$ GeV) where the hydrodynamical behavior is expected
to be dominant. 
Therefore, this deviation from the $KE_T$ scaling in the elliptic and triangular flows could
be explained by 
the mass splitting and mass ordering realized by hydrodynamical 
motion. 
To obtain a conclusive result for the quark number scaling in the collective flow at LHC 
measurement at higher transverse kinetic energy, where the effects of
the recombination mechanism are more clearly observed, is indispensable.  

For the experimental confirmation of the quark number scaling in the elliptic flow, 
no particles are more important than the
$\phi$ meson.  Because it is composed of one quark and one antiquark
and its mass is close to that of the proton, analyses of the elliptic flow of $\phi$ reveal the 
dominant process in the hadronization at the intermediate $P_T$:
recombination mechanism or thermal process as described by
hydrodynamics \cite{Nonaka:2003hx}.  
In addition, because multi-strange hadrons have large mass and small hadronic cross 
sections, they should be less sensitive to hadronic rescattering in the later stage 
of collisions and a good probe of the early state of the collisions such as the partonic
elliptic flow. 
STAR Collaboration measured the $v_2$ of multi-strange 
hadrons ($\phi,\Xi$, and $\Omega$) in Au+Au collisions at
$\sqrt{s_{NN}}=200$ GeV and showed the quark number scaling works 
in the elliptic flow of the multi-strange hadrons \cite{Shi:2011wp}. 
This implies that
the partonic collectivity is built up at the top RHIC energy.  

Furthermore STAR Collaboration investigates the energy dependence of the elliptic flow 
as a function of the transverse momentum for $\phi$ at RHIC\cite{Nasim:2011ju}. 
The quark number scaling in the elliptic flows of $\pi$, $K$, $p$, $\Lambda$, and $\Xi$ 
works well at both $\sqrt{s_{NN}}=39$ and 11.5 GeV in Au+Au collisions. 
They found that while at $\sqrt{s_{NN}}=39$ GeV the elliptic flow of $\phi$  
also follows the quark number scaling, at $\sqrt{s_{NN}}=11.5$ GeV 
the elliptic flow per constituent quark of $\phi$ is much smaller 
than those of the other particles. Only the $\phi$ meson is out of the quark number scaling of 
the elliptic flow at this energy. They conclude that it indicates the dominance 
of hadronic interactions with the decrease of the beam energy. To confirm this,
measurement of the elliptic flow of $\Omega$ would be helpful. 

At the top RHIC energy, the elliptic flow of hadrons which contain only light flavor quarks 
($u,d$, and $s$) follows the quark number scaling. Does the quark number scaling 
work in the elliptic flow of hadrons which are composed of heavy flavors? 
Recent STAR measurement of the elliptic flow of $J/\psi$ gives the answer to this question. 
It shows that the elliptic flow of $J/\psi$ is consistent with zero 
over the measured $P_T$ range, which is significantly smaller than that 
of $\phi$ and inclusive charged hadrons \cite{Tang:2011kr}. 
The picture that $J/\psi$ production is dominated by charm quark recombination with 
significant charm quark flow is disfavored.  On the other
hand, the elliptic flow of open charm mesons is well understood by
the recombination of a light (anti)quark and an (anti)charm quark with
elliptic flow \cite{Laue:2005pw}. 

Next we argue that the recombination model is one of powerful tools to understand 
hadron properties and the QCD phase diagram with high energy heavy ion collisions. 
The investigation of the elliptic flow of hadron resonances with the recombination  
picture at RHIC makes it possible to know final state interactions in the
freezeout processes and the 
structure of resonances \cite{Nonaka:2003ew}. 
In principle, there are two different mechanisms that contribute to resonance production
and as a result there are two types of resonances:
(1) primordial resonances - resonances 
produced from hadronizing quark gluon plasma (QGP mechanism), 
and (2) secondary resonances - resonances produced in the hadronic final state via hadron-hadron 
scattering (HG mechanism).  For example, in the case of $K^*_0$, which is composed of 
the $d$ and $\bar{s}$ quarks, $K^*_0$ are produced by the recombination
of the $d$ and $\bar{s}$ quarks in the QGP mechanism, while it is created
via scattering of $K$ and $\pi$ in the HG mechanism.
The recombination model tells us that the scaling constant $n$ of
the elliptic flow of a hadron species is 
given by the number of constituent quarks which participate in the production of the hadron:
in the case of the QGP mechanism  $n=2$, and
in the case of HG mechanism $n=4$.
From these two contributions the measured elliptic flow is given by, 
\begin{equation}
v_2^{\rm measured} = r(P_T)v_2^{\rm QGP}+ (1-r(P_T))v_2^{\rm HG},
\end{equation}
where $r(P_T)$ is the fraction of resonances which were produced at the hadronization
and whose decay products escaped from the hadron phase
without rescattering.
With $r(P_T)$ one can investigate the cross sections of resonances in the hadronic medium.  
STAR attempted to extract the scaling constants (or apparent constituent quark number)
from the measured elliptic flow of $K^{0*}$ to investigate its 
production mechanism. Unfortunately, the obtained value of $n$ is $n=3\pm2$ \cite{Adams:2004ep},
from which we cannot draw a definite conclusion. 
This method is easily extended to explore the structure of exotic hadrons, such as tetraquark and 
pentaquark.  Utilizing the scaling law of the elliptic flow, one can obtain the information on
the structure of exotic hadrons, whether they have
compact structure or are more like molecular bound states \cite{Nonaka:2003ew}.  

We can explore consequences of diquark and quark-antiquark clustering above 
the deconfinement temperature with event-by-event net charge fluctuations 
\cite{Nonaka:2005vr}.
Recently, lattice QCD calculations show that charmonia survive even well above the critical 
temperature \cite{Asakawa:2003re,Datta:2003ww,Iida:2006mv,Umeda:2007hy},
which suggests the possibility that hadronization occurs via recombination
of not only single $q$ or 
$\bar{q}$ but also $qq$ or $q\bar{q}$. 
The $D$ measure, the net charge fluctuation normalized by the entropy is defined by
\begin{equation}
D=4\langle (\Delta Q)^2\rangle/N_{\rm ch}, 
\end{equation}
where $\langle (\Delta Q)^2\rangle$ denotes the event-by-event net 
charge fluctuation within a given rapidity window $\Delta y$ 
and $N_{\rm ch}$ is the total number of charged particles emitted in 
this window. For a free plasma of quarks and gluons $D\approx 1$, 
whereas for a free pion gas $D \approx 4$. However, experimental data 
at RHIC is rather close to the value of hadron gas $D=2.8 \pm 0.05$ in 
central Au+Au collisions at $\sqrt{s_{NN}}=130$ GeV (STAR) \cite{D_STAR} 
and $D\approx 3$ (PHENIX) \cite{D_PHENIX}. 
In the recombination scenario, the fluctuation of 
the net charge $Q=\sum_iq_i n_i$ is given as 
\begin{equation}
\langle \delta Q^2 \rangle \equiv \langle Q^2 \rangle - \langle Q \rangle ^2 
= \sum_i(q_i)^2 \langle n_i \rangle + \sum_{i,k}c^{(2)}_{ik} \langle n_i \rangle 
\langle n_k \rangle q_i q_k, 
\label{eq:delta_Q}
\end{equation}
where $c^{(2)}_{ik}$ are the normalized two-particle correlation functions. 
In the absence of two-particle correlations, Eq.~(\ref{eq:delta_Q}) can be 
rewritten as $\langle \delta Q^2\rangle = \frac{4}{9} (N_u + N_{\bar u}) + 
\frac{1}{9}(N_d + N_{\bar d} + N_s + N_{\bar s} ) $, where 
$N_i=\langle n_i \rangle$ denotes the average number of the constituent 
quarks with flavor $i$. 
Together with the statistical hadronization model and the multiplicity of experimental value, 
the net charge fluctuation in the quark recombination scenario is 
$d\langle \delta Q^2 \rangle_{\rm q}/dy=331\pm27$, which is close to 
the experimental value $d\langle \delta Q^2 \rangle_{\rm had}/dy=368\pm33$ \cite{D_STAR}. 
Furthermore the difference of the net charge fluctuations between the quark recombination and 
the experimental data suggests necessity of improvement on both theory and experiment sides.
One of such possibilities is to include $qq$ and $q\bar{q}$ pairs to the hadronization 
mechanism \cite{Nonaka:2005vr}.  If the $qq$ pair and $q\bar{q}$ are taken into account, 
Eq.~(\ref{eq:delta_Q}) is extended to  
\begin{equation}
\langle \delta Q^2 \rangle = 
\sum_i (q_i)^2 (N_i + N_{\bar{i}}) 
+ \sum_{ij}(q_i + q_j)^2 \langle n_{ij}\rangle 
+ \sum_{ij} (q_i - q_j)^2 \langle \bar{n}_{ij}\rangle\, , 
\end{equation}
where the average numbers of diquarks and $q\bar{q}$ are proportional to the products of the 
individual quark numbers: $\langle n_{ij}\rangle=\alpha (N_i N_j + N_{\bar{i}}N_{\bar{j}})$, 
$\langle \bar{n}_{ij}\rangle\ = \beta N_iN_{\bar{j}}$ with relative pairing weights $\alpha$ and $\beta$.  
They showed that experimental data can be fitted with an appropriate choice of
$\alpha$ and $\beta$.

In Ref.\cite{Nonaka:2005vr}, the problem of the gluonic degrees of freedom
was also discussed. According to recent lattice calculations\cite{Borsanyi:2010cj, Bazavov:2011nk,Umeda:2010ye},
the phase transition at small baryon chemical potential is crossover.
This means that the hadronization at RHIC and LHC does not
take place from a system where quarks and gluons are active degrees of
freedom, but from a state with reduced entropy density. At the moment, 
no method on the lattice has been found to figure out what is the
active degrees of freedom in the crossover region. In Ref.~\cite{Nonaka:2005vr}
it was shown that entropy density data on the lattice are not
inconsistent with that of a gas of constituent quarks, 
which is assumed in the recombination model. The success
of the recombination model strongly suggests that the gluonic
degrees of freedom disappear first when the temperature is
decreased from well-above the (pseudo)critical
temperature. However, it is needless to say that it is important
to find a way to figure out how the active degrees of
freedom change across the crossover region on the lattice
in order to steady this picture.

\section{The Freezeout Process \label{sec:freezeout}}
As fireballs expand, the temperature and density inside them become small.  
Finally the mean free path among particles inside the fireballs becomes so large that
the assumption for the hydrodynamical picture becomes invalid. 
Currently two separate freezeout processes are believed to occur successively in  
heavy ion collisions. One is chemical freezeout, at which the ratios 
of hadrons are fixed, and the other is thermal (kinetic) freezeout, at which 
the particles stop interacting. Recently final state interactions between the two freezeout
processes are also included in hydrodynamical models by connecting
hadron based event generators to the hydrodynamical calculations.
In this picture, kinetic freezeout is found to be not an instantaneous
process but a continuous one as we show in the following.

\subsection{Chemical Freezeout and Thermal Freezeout}
The chemical freezeout temperature and baryon chemical potential are extracted with  
the statistical model on the basis of the grand canonical formalism. 
Surprisingly, statistical models are in excellent agreement with experimental 
data for hadron ratios in a wide range of collision energy from SIS to 
RHIC \cite{rev-particle-production}. 
At present the (pseudo)critical temperature suggested 
from the latest lattice calculation is $T_{\rm c}= 150 \sim 160$ MeV (see Subsec.\ \ref{subsec:EoS}), 
which is relatively low from the point of view of the statistical model \cite{Andronic:2011yq}. 
For example, using a statistical model, Andronic et al. extract the chemical freezeout temperature 
$T_{\rm ch}=160 \sim 166$ MeV for LHC Pb+Pb collisions at $\sqrt{s_{\rm NN}}=2.76 $ GeV.
It might be too early to take lattice results at the face value, but this suggests that 
the physical meaning of the temperature and chemical potential obtained by the statistical 
model need to be reexamined. 
At the same time, one needs to make it clear why the statistical model works 
very well not only in a wide range of collision energies from SIS to LHC but also 
for smaller systems such as $p$+$p$ collisions \cite{Becattini:2009sc}.

At the thermal freezeout temperature, the mean free paths of the hadrons
have grown to the order of the system size and hydrodynamical 
description, which requires very small mean-free path, is clearly no longer
applicable. A first naive guess for the kinetic freezeout temperature
$T_{\rm f}$ would be of the order of the pion mass ($\sim140$ MeV). Practically the value of 
the kinetic freezeout temperature in hydrodynamical models is determined 
from comparison with data of the $P_T$-spectra.

The task at the end of a hydrodynamical calculation is to populate fluid cells
with particles with the final temperature and flow.
For calculations of single particle spectra, the simple assumption of
a sudden freezeout process at a certain proper time for each fluid cell 
is often adopted, neglecting the reverse  process from particles to the
hydrodynamical medium.
Under this assumption the Cooper-Frye formula \cite{CoFr74} is widely used, 
\begin{equation}
E\frac{dN}{d^3P} =  
\sum_h\frac{g_h}{(2\pi)^3} \int_\Sigma d\sigma_\mu 
P^\mu \frac{1}{\exp[(P_\nu u^\nu-\mu_f)/T_{\rm f}]\pm 1}, 
\label{Eq-CF}
\end{equation}
where $g_h$ is the degeneracy factor of hadron $h$ and $T_{\rm f}$ and $\mu_{\rm f}$ are 
the freezeout temperature and baryon chemical potential, respectively, and
$+(-)$ is for fermions (bosons). 
$d\sigma_\mu$ is obtained by calculating
the normal vector on the freezeout hypersurface $\Sigma$. 
Once these quantities are given, using Eq.\ (\ref{Eq-CF}),
one can calculate the distribution of any particle
after freezeout.

More realistic models have been 
investigated. One of them is the Continuous Emission Model (CEM),
in which particles are emitted continuously from the whole expanding 
volume of the system at different temperatures and different times
\cite{Socolowski:2004hw}.  
In the early days of hydrodynamics only kinetic freezeout was implemented.
Indeed, at lower collision energies such as at SIS and AGS, the separation 
between the chemical freezeout and kinetic freezeout points is not large
on the $T-\mu$ plane.
However, at RHIC there is a significant difference between kinetic freezeout 
temperatures from hydro-inspired models and the chemical freezeout temperature 
from the statistical model. \cite{Adams:2003xp}. 
This fact also manifests itself through the failure to get the
correct absolute normalization of some $P_T$-spectra, e.g., 
that of protons
in hydrodynamical calculations, with only a kinetic freezeout 
\cite{Heinz:2001xi}. 
For these reasons, attempts to model separate freezeout processes 
consistently via modified equations of state were started.
\cite{Hirano:2002ds,ArGrHaSo01,Arbex:2001vx,Te02}. 

\subsection{Final State Interactions}
\begin{figure}
\centerline{\includegraphics[width=13.5 cm]{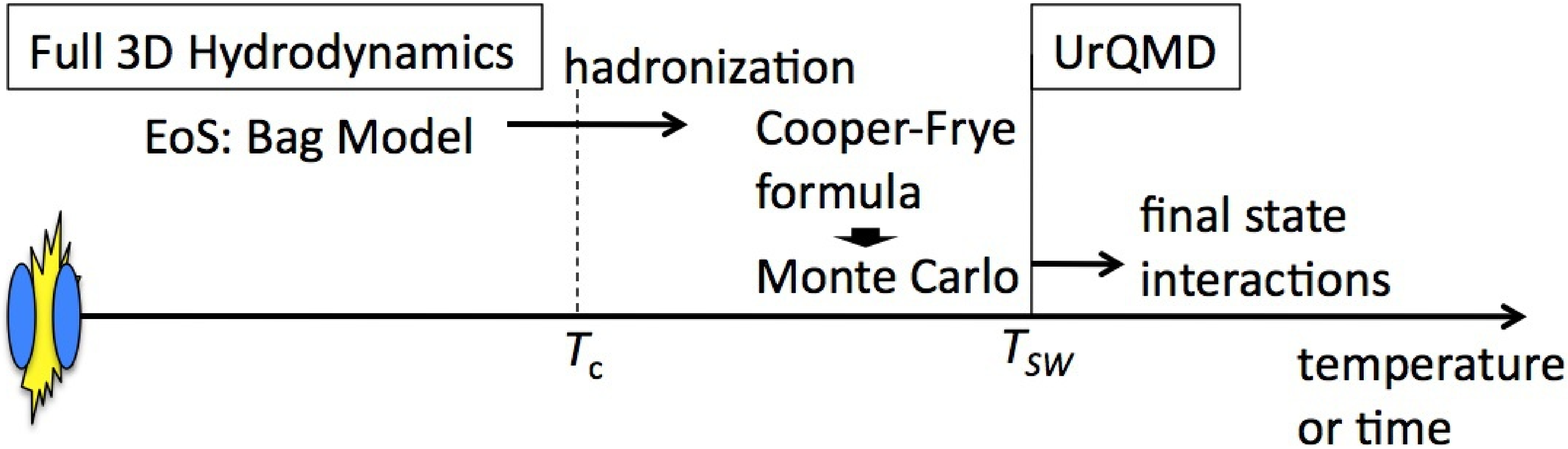}}
\caption{Schematic sketch of the 3D hydro+UrQMD model.	
$T_{\rm c}$ and and $T_{\rm SW}$ are taken to be 160 MeV and
150 MeV, respectively. 
}
\label{fig:hydro_urqmd}
\end{figure}

It turns out that some experimental data are still not understood in 
a satisfactory way even with the two separate freezeouts. 
For example, mean transverse momentum $\langle P_T \rangle$ as a 
function of particle mass does deviate from the linear scaling law, which 
suggests significant final state interactions in the hadronic phase 
\cite{Nonaka:2006yn}. To explain these effects, and to account for the
apparently large viscosities in the hadronic phase, as discussed before,
hydro+cascade hybrid models were introduced.
They use a hydrodynamical computation of the expansion and cooling of hot 
QCD bulk matter, and then couple the output consistently to a hadron-based 
transport model in order to take account of the final state interactions.
A pioneering work on hydro+cascade hybrid models was done by Bass et al.\
\cite{Bass:2000ib} using UrQMD. Similar investigations were carried out 
in Refs.\ \cite{Teaney:2000cw, Hirano:2005xf}.
Figure \ref{fig:hydro_urqmd} shows a schematic sketch of the 
hydro + UrQMD model \cite{Nonaka:2006yn}.
At the switching temperature $T_{\rm SW}$, the mean free path 
of hadrons becomes so large that the hydrodynamical picture becomes 
inapplicable. Thus after this point,
the motion of hadrons is described by UrQMD. 
In practice, at $T_{\rm SW}$ hadron distributions are calculated
with the Cooper-Frye formula in the fluid and 
the initial state of UrQMD is produced with them through
the Monte Carlo methods. 
Then the UrQMD simulation is started. 
The reverse process from UrQMD to fluid dynamics
is neglected in the simulation.

\subsection{Experimental Data and Discussion}
%
The main purpose of our hydrodynamical model + hadron based event 
generator is to handle realistic final state interactions in the
freezeout process. We need to choose an appropriate hadron based
event generator for it. Hadron based event generators have been
developed to understand experimental data especially
at lower energy heavy ion collisions.
Here we use the UrQMD model for the description of final state
interactions in the hadron phase. It gives us reasonable hadron
yields, single particle spectra and so on, and it is often used
for understanding the baseline of experimental data of
relativistic heavy ion collisions.

Here we make a comment on the applicability of cascade models
for the description of the afterburner of hydrodynamical
models. Recent lattice calculations (Ref. \cite{Borsanyi:2010cj}) suggest
that the hadron phase is well-described as a hadron resonance
gas up to the vicinity of the phase transition. Thus, cascade
models are, in general, ideal machinery to investigate its
time evolution and freezeout process.

If the final state interactions are implemented properly, hydrodynamical models 
acquire more predictive power for experimental observables. 
We will show this by comparing single particle spectra and elliptic flows
with experimental data. 
In this subsection we show results calculated with the same 
initial conditions and the equation of state as those in pure hydrodynamical calculations: 
Glauber type initial conditions and a bag model equation of state. 
In this case the maximum value of initial energy density is 40 GeV/fm$^3$, which 
is smaller than that of the pure hydrodynamical model, because the inclusion of
resonance decays and final state interactions to hydrodynamical models 
increases particle multiplicity.  

\begin{figure}
\begin{minipage}[h]{7cm}
\vspace{3mm}
\centerline{\includegraphics[width=7 cm]{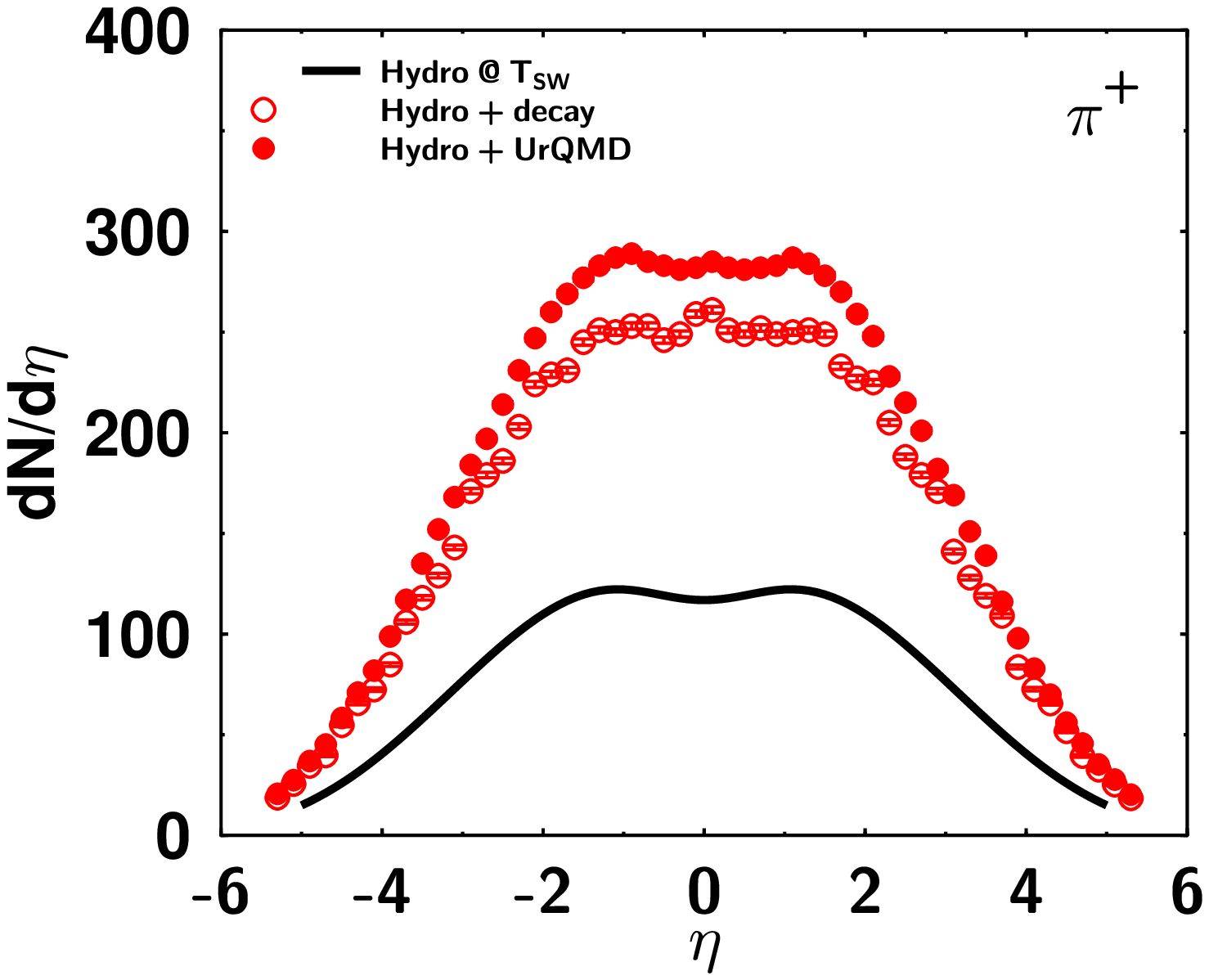}}
\caption{Pseudo rapidity distribution of $\pi^+$ from hydro 
at $T_{\rm sw}$ (solid line), hydro + decay (open circle), and 
hydro + UrQMD (solid circle) in Au+Au $\sqrt{s_{NN}}=200$ GeV
central collisions. 
}
\label{fig:pi_eta}
\end{minipage}
\hspace{1cm}
\begin{minipage}[h]{7cm}
\centerline{\includegraphics[width=7.0 cm]{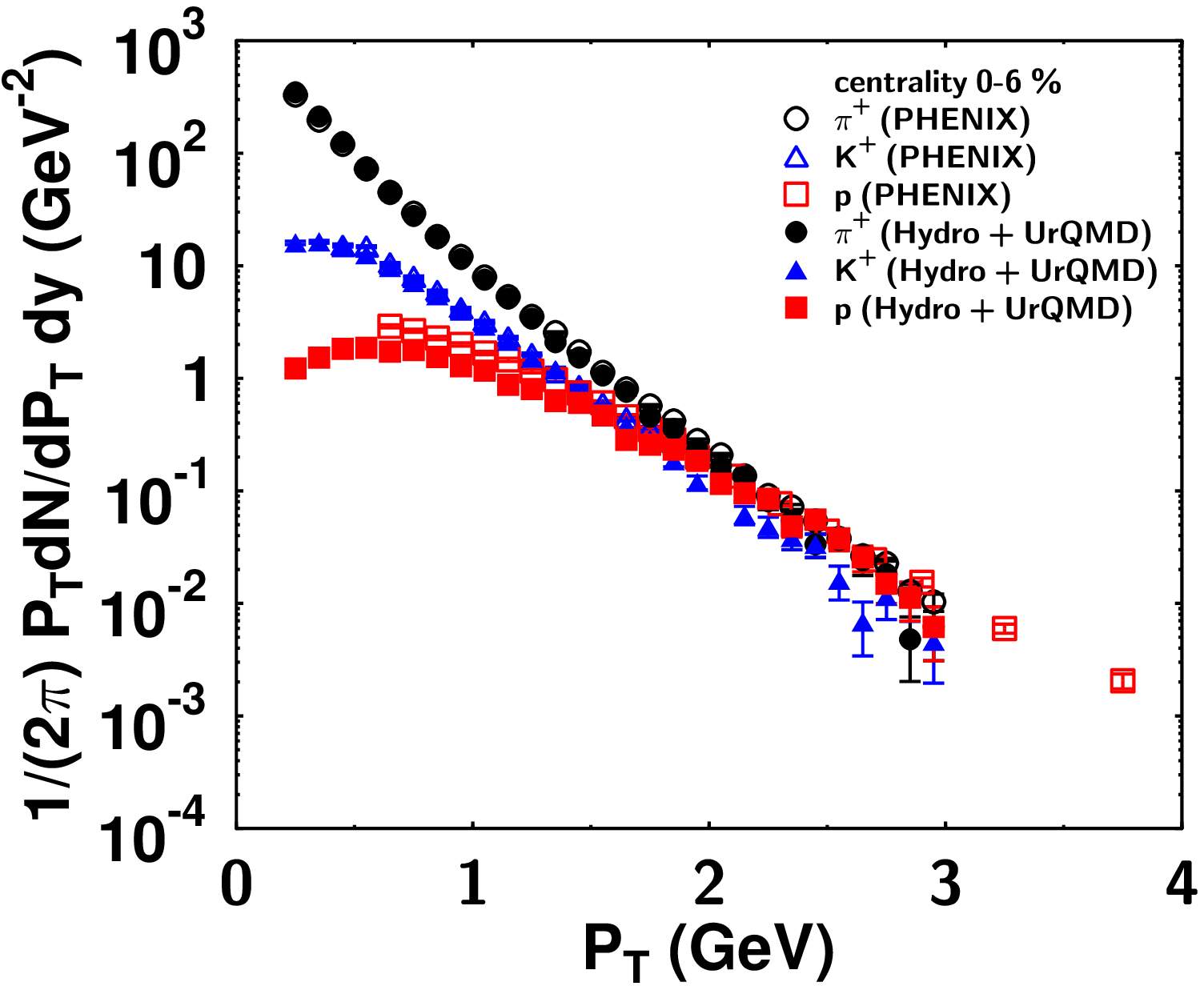}}
\caption{Comparison of the calculated $P_T$ spectra for $\pi^+$, $K^+$, and $p$
in central collisions with PHENIX data \cite{PHENIX_pt}. The points are not renormalized. } 
\label{fig:hu_pt_b24}
\end{minipage}
\end{figure}

First let us consider the argument on the multiplicity, which
we mentioned in Sec.\ \ref{sec:IC}. 
We take $\pi^+$ as an example, 
because it is one of the most dominant particles
in the charged hadron multiplicity and 
affected a lot from resonance decays.
Figure \ref{fig:pi_eta} shows the pseudo rapidity distributions of 
$\pi^+$ from the fluid at $T_{\rm sw}$ (solid line), 
that from the fluid plus resonance decay effects (open circles), 
and its final state multiplicity (solid circles). 
We find that the resonance decay effect is huge and that the multiplicity 
including resonance decay effect (open circles) becomes almost twice 
as many as that of hydro at $T_{\rm sw}$ (solid line).  
Furthermore, the additional gain from final state interactions
is seen. This result clearly shows that the dynamics of the hadron
phase such as resonance decays  and final state interactions,
is important also in high energy heavy ion collisions. 
In short, final state multiplicities cannot be predicted
by determining only the initial state.

Figure \ref{fig:hu_pt_b24} shows the $P_T$ spectra of $\pi^+$, $K$, and $p$ in 
central collisions at $\sqrt{s_{NN}}=200$ GeV. The most remarkable feature,
compared to the pure hydrodynamical calculation, is that the hydro+micro 
approach is capable of accounting for the proper normalization of
the spectra for all hadron species
without any additional  
correction as is performed in the pure hydrodynamical model. 
The introduction of realistic freezeout process  
provides therefore 
a natural solution to the problem of the separation of the chemical and 
kinetic freezeouts in pure hydrodynamical calculation. Similar results
have been obtained previously in 1+1 and 2+1 dimensional implementations
of the hydro+micro approach \cite{BaDu00,Teaney:2001av}.

\begin{figure}
\centerline{\includegraphics[width=7.5 cm]{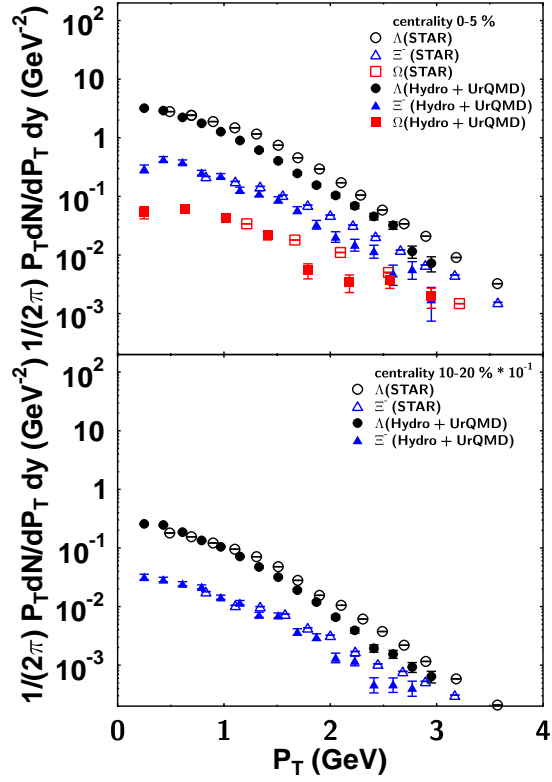}}
\caption{Calculated $P_T$ spectra of multi-strange particles at centralities 0--5 \%  
and 10--20 \% and STAR data \cite{STAR_strange1}. 
} 
\label{fig:pts}
\end{figure}

In Fig.~\ref{fig:pts} we analyze the $P_T$ spectra of 
multi-strange particles. Our results show good agreement with   
experimental data for $\Lambda$, $\Xi$, and $\Omega$ for centralities 0--5 \% 
and 10--20 \%.  In this calculation no additional procedure for normalization 
is needed.  
Recent experimental results suggest that 
at the thermal freezeout multi-strange baryons exhibit less transverse flows  
and higher temperatures closer to the chemical freezeout 
temperature compared to non- or single-strange baryons
\cite{STAR_strange1,STAR_strange2}. This behavior can be understood in terms
of the flavor dependence of the hadronic cross section, which decreases
with increasing strangeness content of the hadron. The reduced
cross section of multi-strange baryons leads to a decoupling from the
hadronic medium at an earlier stage of the reaction, allowing them
to provide information on the properties of the hadronizing QGP less
distorted by hadronic final state interactions 
\cite{vanHecke:1998yu,Dumitru:1999sf,Teaney:2001av}.
It should be noted that the analogous behavior has already been observed
in experiments at CERN SPS \cite{sps_multistrange}.

\begin{figure}
\hspace{-0.5cm}
\begin{minipage}[h]{7cm}
\centerline{\includegraphics[width=7 cm]{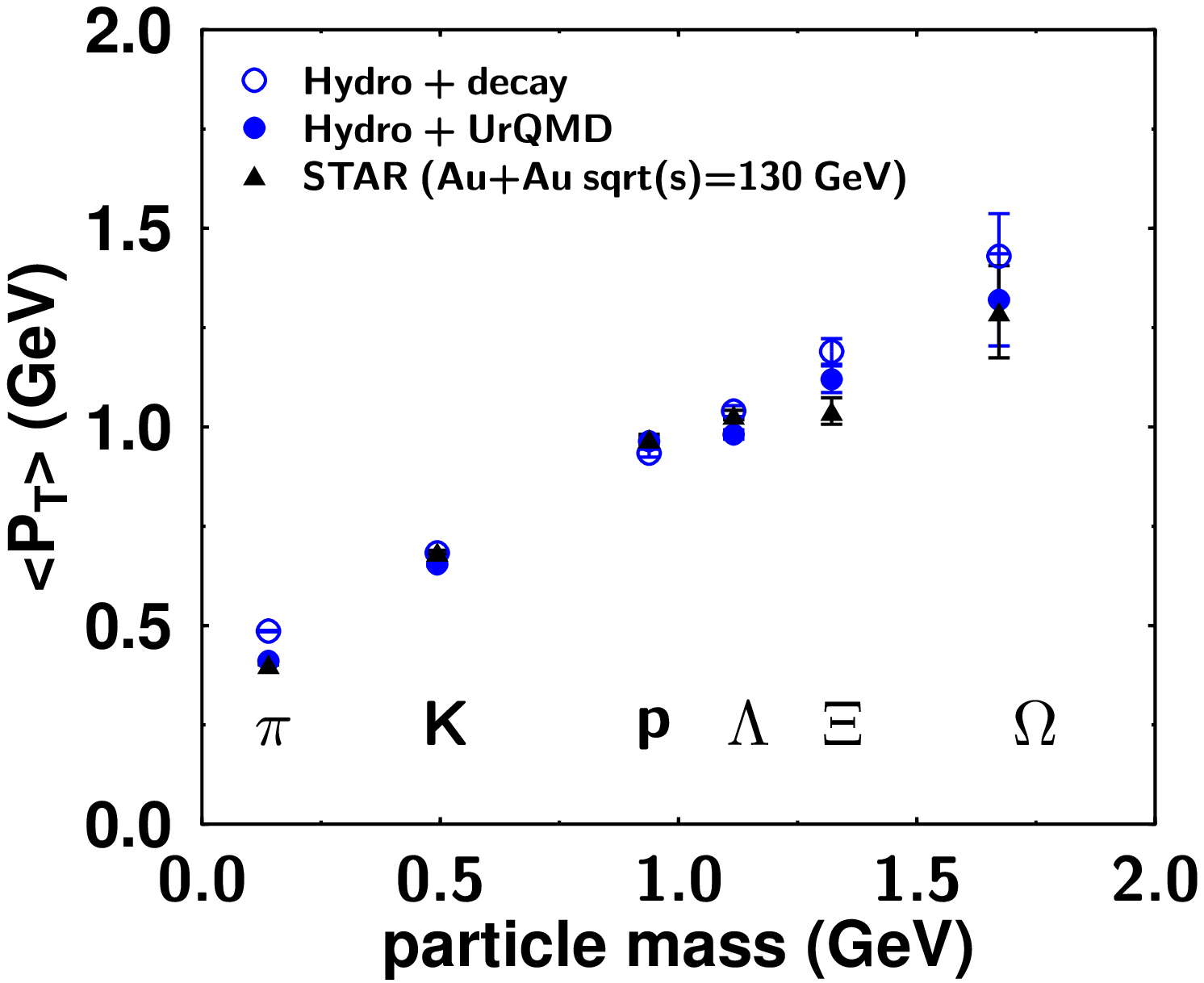}}
\caption{Mean $P_T$ as a function of mass. The open circle symbols stand for 
hydro + decay, solid circle symbols stand for hydro + UrQMD, and solid 
triangle symbols stand for STAR data (Au+Au $\sqrt{s_{NN}}=130$ GeV).}
\label{fig:meanpt}
\end{minipage}
\hspace{0.1cm}
\begin{minipage}[h]{7cm}
\vspace{-0.4cm}
\centerline{\includegraphics[width=7.0cm]{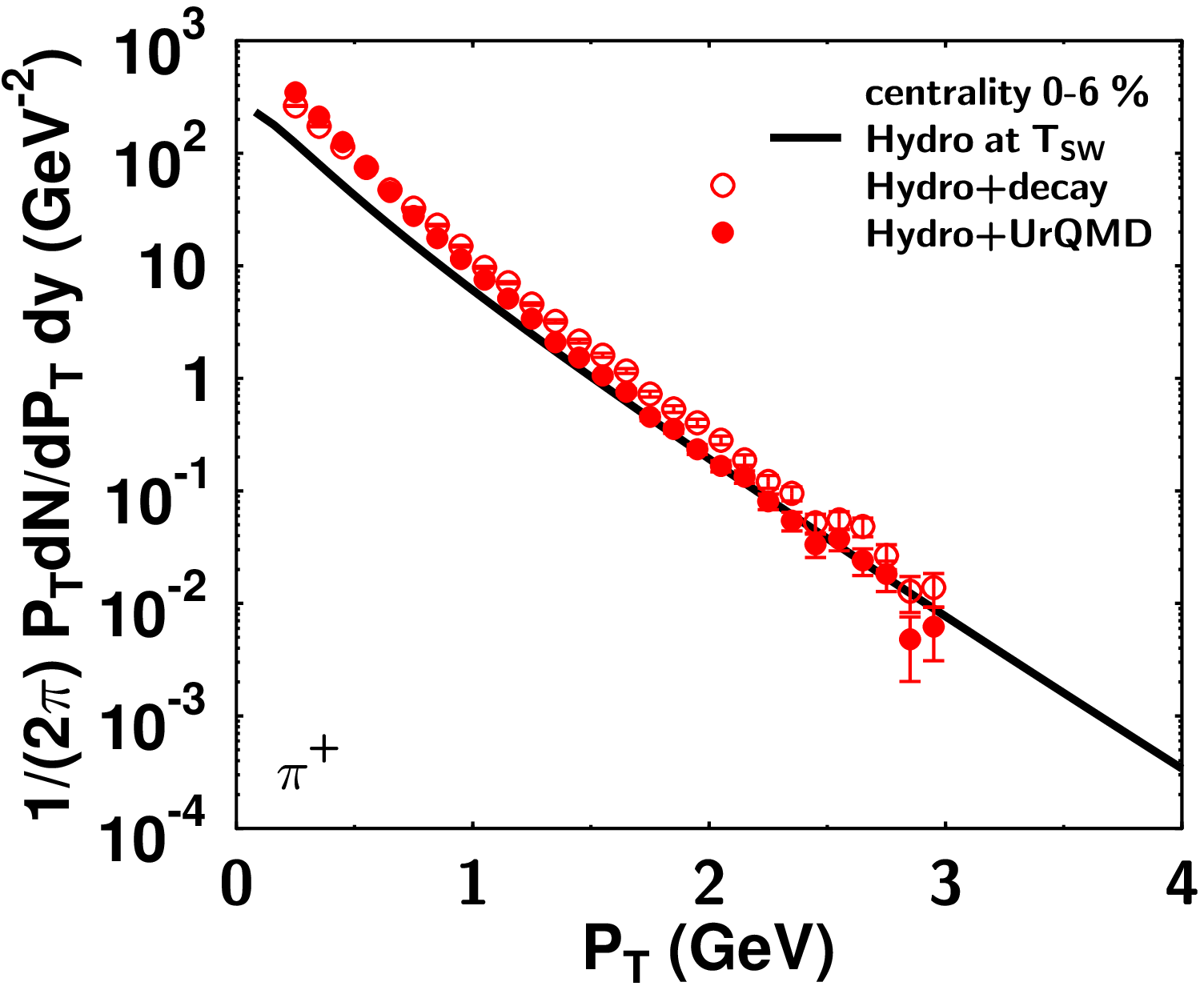}}
\caption{$P_T$ spectra of $\pi^+$ from hydro at the switching 
temperature (solid line), hydro+decay (open symbols) and 
hydro + UrQMD (solid symbols) in central collisions.}
\label{fig:pt_decay}
\end{minipage}
\end{figure}

In Fig.~\ref{fig:meanpt} the mean transverse momentum $\langle P_T\rangle $ 
as a function of the hadron  
mass is shown. Open symbols denote the value at $T_{\rm sw}=150$~MeV, corrected
for hadronic decays. Not surprisingly, in 
this case the $\langle P_T\rangle $ follows a straight line, suggesting a 
hydrodynamical expansion.  However, if hadronic rescattering 
is taken into account (solid circles), the $\langle P_T\rangle $  
does not follow the straight line any more.
The $\langle P_T\rangle $ of pions is actually reduced by hadronic
rescattering (they act as a heat-bath in the collective expansion), 
whereas protons actually pick up additional transverse momentum in the
hadronic phase. RHIC data by STAR collaboration are shown by
the solid triangles -- overall, the proper treatment of hadronic
final state interactions significantly improves the agreement of the
model calculation with the data.
We note that our results confirm those
previously obtained in 1+1 and 2+1 dimensional implementations
of the hydro+micro approach  \cite{BaDu00,Teaney:2001av}, demonstrating
the robustness of the hydro+micro approach across the three different
implementations of the hydrodynamical and macro to micro transition components
of the model. 

Let us now investigate the effect of resonance decays and
hadronic rescattering on the pion and
baryon transverse momentum spectra. Figure~\ref{fig:pt_decay}
shows the $P_T$ spectrum of $\pi^+$ at $T_{\rm sw}=150$~MeV
(solid line, uncorrected for resonance decays) 
as well as the final spectrum after hadronic rescattering and
resonance decays, labeled as 
{\em Hydro+UrQMD}
(solid symbols). 
In addition, the open symbols denote a calculation
with the resonance decay correction
performed at $T_{\rm sw}$, which we label as {\em Hydro+decay}.
The difference between the solid line and open symbols therefore
directly quantifies the effect of resonance decays on the spectrum,
which is most dominant in the low transverse 
momentum region $P_T < 1$ GeV.
Furthermore, the comparison 
between the open symbols and solid symbols quantifies the effect of hadronic
rescattering:
pions with $P_T > 1 $ GeV lose momenta via these final state 
interactions, resulting in the steeper slope.

\begin{figure}
\hspace{-0.5cm}
\begin{minipage}[h]{7cm}
\centerline{\includegraphics[width=7 cm]{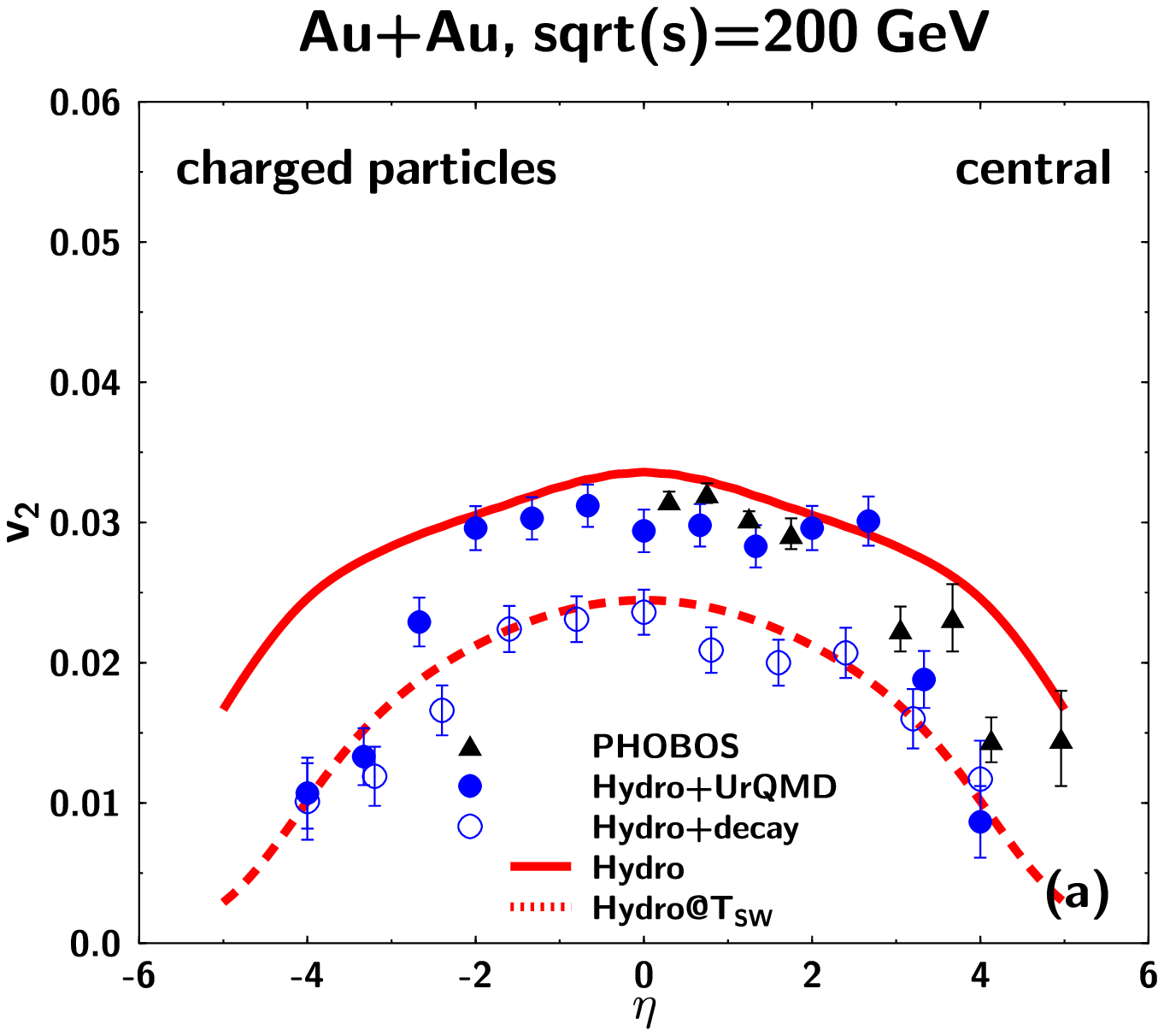}}
\caption{Elliptic flow as a function of $\eta$  for $\pi^+$ at 
centralities 5-10 \% and 10-20 \%. The open symbols stand for 
STAR data and solid symbols stands for our results. 
}
\label{fig:v2_eta}
\end{minipage}
\hspace{0.1cm}
\begin{minipage}[h]{7cm}
\vspace{-0.4cm}
\centerline{\includegraphics[width=7.0cm]{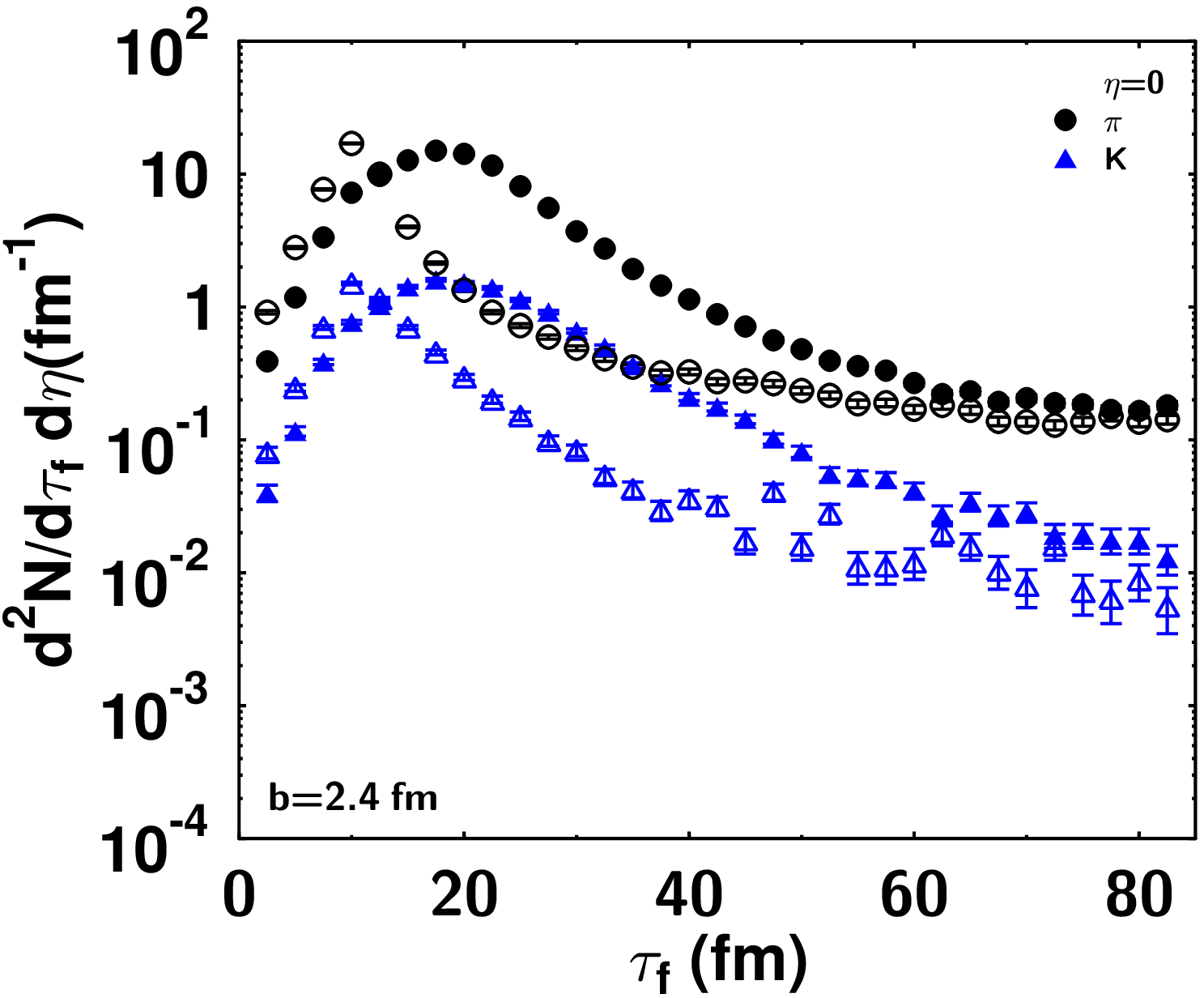}}
\caption{Freezeout time distribution of mesons for hydro+decay (open 
symbols) and hydro + UrQMD (solid symbols) at the mid-rapidity in the 
case of central collisions.}
\label{fig:ntau_m}
\end{minipage}
\end{figure}

Figure \ref{fig:v2_eta} shows the elliptic flow as a 
function of $\eta$: the pure hydrodynamical calculation is shown by 
the solid curve, the hydrodynamical contribution at $T_{\rm sw}$ is denoted
by the dashed line and the full hydro+micro calculation is given by
the solid circles, together with PHOBOS data (solid triangles). 
The shape of the elliptic flow in the pure hydrodynamical calculation at 
$T_{\rm sw}$ is quite different from that of the full
hydrodynamical one terminated at the kinetic freezeout temperature, 110~MeV. 
Apparently the slight bumps at forward and backward rapidities observed
in the full hydrodynamical calculation develop in the later hadronic
phase, since it is not observed in the calculation terminated at 
$T_{\rm sw}$.
Evolving the hadronic phase in the hydro+micro approach increases
the elliptic flow at the central rapidities, but not in the projectile and 
target rapidity regions.
As a result, the result for the elliptic flow in the hydro+micro approach
is closer to the experimental data, compared to the pure hydrodynamical calculation.

In Fig.~\ref{fig:ntau_m} we investigate the effect of hadronic rescattering
on the duration of the freezeout process by
comparing the calculation terminated
at $T_{\rm sw}$ without hadronic rescattering (open symbols) and the
one including the full hadronic final state interactions (solid symbols). 
If we terminate time evolution at $T_{\rm sw}$, most hadrons  
freezeout around $\tau_f =$ 10 fm/c, reflecting the lifetime of the deconfined phase
(the tails of the distribution come from the decays
of long-lived resonances).
The inclusion of hadronic rescattering shifts
the peak of the freezeout distribution to larger freezeout 
times ($\tau_f \sim 20-30$ fm), which provides an estimate on the
lifetime of the hadronic phase around 10-20 fm. 

\begin{figure}
\centerline{\includegraphics[width=7.5cm]{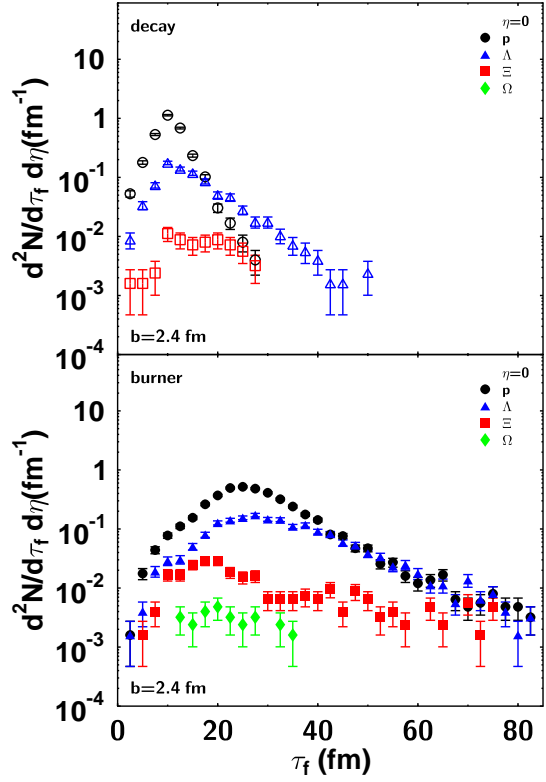}}
\caption{Freezeout time distribution of baryons for 
hydro+decay (open symbols, above) and hydro + UrQMD (solid symbols, below) 
at the mid-rapidity.
} 
\label{fig:ntau_b}
\end{figure}
The findings discussed in the context of the previous figure for
pions and kaons are confirmed by analyzing baryons in the same way,
which is shown in Fig.~\ref{fig:ntau_b}: here the top frame
contains the analysis terminated at $T_{\rm sw}$ and the bottom frame
contains the calculation including full hadronic rescattering. 
Figures \ref{fig:ntau_m} and \ref{fig:ntau_b} suggest that
the assumption of sudden freezeout, which is often used, for example,
in the blast wave model,
is not realistic under the existence of final state interactions.  

\section{Summary \label{sec:sum}}

In this article, we outlined key issues in constructing a realistic
and comprehensive dynamical model 
for the description of the whole stages of relativistic high energy heavy
ion collisions:
initial conditions, hydrodynamical expansion, hadronization, and freezeout processes.   
Hydrodynamical models are a promising starting point for building
multi-module models. One of the reasons is that they give global 
understanding of experimental data at RHIC and their success at RHIC makes us to 
expect that hydrodynamical analyses will serve as the baseline for
investigation also at LHC. 
In addition, one can easily input up-to-date knowledge on each stage in
the time evolution to hydrodynamical
models easily: 
more realistic initial conditions which contain even-by-event fluctuations, 
latest equations of state and transport coefficients from lattice QCD, 
realistic hadronization 
mechanisms such as those from the recombination model and fragmentation mechanism,
and realistic freezeout processes including final state interactions.  

Presently, both ideal and viscous hydrodynamical models are utilized for
the investigation of data obtained at RHIC and LHC.
In the multi-module modeling, however, ideal hydrodynamical 
models have been studied more deeply and its status is considered to be 
more mature than viscous hydrodynamical models. 
However, gradually the majority of hydrodynamical models
will become viscous hydrodynamical models instead of ideal hydrodynamical models,
because the study of viscosity effects in
physical observables and the estimate of the viscosities in QGP
from hadronic observables are among the hottest topics in the field;
inclusion of viscosities in hydrodynamics is inevitable for such investigation. 
At the moment, investigation with viscous hydrodynamical models has just started. 
Whereas their development is fast and remarkable, a lot of issues remain to be 
considered, implemented, tested, and improved. 

One of them is numerical schemes for solving the relativistic hydrodynamical
equation, to which only a little attention has been paid up to now.
In particular, the estimate of numerical viscosity is crucial
for the study of the viscosities of the matter
created in relativistic heavy ion collisions.
In Sec.~\ref{sec:NS}, we showed that SHASTA, KT, and rHLLE schemes, which 
are mainly used in studies of high energy heavy ion collisions, have nearly 
the same accuracy and numerical artifact. 
Furthermore, we showed the results of shock tube test with a new numerical 
scheme which suffers less artificial dissipative effect and showed that
it is more suitable for analyses of physical viscosities
than SHASTA, KT, and rHLLE schemes. 

In summary, in this article we discussed essential ingredients in understanding 
entire stages of high energy heavy ion collisions in detail, together with 
interpretation of experimental data at SPS, RHIC, and LHC. 
We hope that the road map shown in this article will serve as a guideline
for modeling a realistic dynamical model for high energy heavy ion collisions. 

\section*{Acknowledgments}
We would like to thank Harry Niemi for providing us with their results 
shown in Figs.~\ref{fig:ene_ns} and \ref{fig:vx_ns}. 
This work was in part supported by Grant-in-Aid for Young Scientists (B) (22740156) 
and Grant-in-Aid for Scientific Research (S)(22224003) and (C)(23540307).

\end{document}